\documentclass[%
 aip,
 amsmath,amssymb,
 preprint,%
]{revtex4-1}

\usepackage{graphicx}
\usepackage{dcolumn}
\usepackage{bm}

\usepackage{graphicx,epic,amscd,amssymb,amsmath,multirow}
\usepackage[all]{xy}
\usepackage{subfigure}
\usepackage{pict2e}

\newcommand{\qed}{\hfill $\Box$}
\def\C{{\mathbb C}}

\def\R{{\mathbb R}}

\newtheorem{theorem}{Theorem}
\newtheorem{corollary}{Corollary}
\newtheorem{remark}{Remark}

\newtheorem{lemma}{Lemma}
\newtheorem{proposition}{Proposition}

\begin{document}
\renewcommand{\theenumi}{(\arabic{enumi})}
\renewcommand{\labelenumi}{\theenumi}

\preprint{
}

\title[]{Exact solutions to SIR epidemic models via integrable discretization}

\author{Atsushi Nobe}
\email{nobe@waseda.jp}
\affiliation{ 
Faculty of Political Science and Economics, Waseda University,\\ 1-6-1 Nishiwaseda, Shinjuku, Tokyo 169-8050, Japan
}%

\date{\today}

\begin{abstract}
An integrable discretization of the SIR model with vaccination is proposed.
The conserved quantities of the continuous model are inherited to the discrete model through the discretization, since the discretization is based on the intersection structure of the non-algebraic invariant curve defined by the conserved quantities.
Uniqueness of the forward/backward evolution of the discrete model is demonstrated in terms of the single-valuedness of the Lambert W function on the positive real axis.
Furthermore, the exact solution to the continuous SIR model with vaccination is constructed via the integrable discretization.
The discretization procedure similarly applied to the original SIR model leads to two kinds of integrable discretization, and the exact solution to the continuous SIR model is also deduced. 
It is moreover shown that the discrete SIR model geometrically linearizes the time evolution by using the non-autonomous parallel translation of the line intersecting the invariant curve.
%
\end{abstract}

\pacs{02.30.Hq, 02.30.Ik, 05.45.Yv}
\keywords{SIR model, integrable system, conserved quantity, discretization, Lambert W function}
\maketitle

\section{Introduction}
\label{sec:intro}

The SIR model \cite{KM27} is a three-dimensional nonlinear dynamical system firstly proposed in 1927 by Kermack and McKendrick in order to investigate mathematical theory of infectious diseases.
The SIR model provides the differential relation with respect to time among the population densities of the susceptible, the infected and the recovered relevant to the diseases, which is controlled by the transmission rate and the recovery rate.
The most important mathematical feature of the SIR model is that it has two functionally independent conserved quantities; the first one is trivial while the second is nontrivial.
The first trivial one is the total population density immediate from the structure of  the model.
The second one is presented by a kind of logarithmic function in the population densities, and hence is non-algebraic.
Since the three-dimensional SIR model has two independent conserved quantities, the model is considered to be integrable in the sense of Liouville. 
Investigations of the SIR model and its many variants, collectively called the SIR models, have intermittently been done until recently, and their mathematical structures have gradually been revealed \cite{Hethcote74,Dietz75,Schenzle84,WGCR03,SWRGC04,KMVL09,SKS10,HLM14}.
Meanwhile, the pandemic of the COVID-19 since 2020 has drastically boosted the studies on mathematical models relevant to epidemics of infectious diseases \cite{ADLMVV20,KNRR21,Kubota21,GMV21,KO22,TM22}.
In these studies, the SIR models have still played basic and important roles due to their mathematically good lineage and manageability which are considered to be arising from their integrability.

In the studies of mathematical theory of infectious diseases, one often prefer discrete models to continuous ones.
The reason is that discrete models perfectly fit into computer simulation; with actual data of infectious diseases, we can directly compare the exact solutions to the discrete models obtained by mere substitution of the initial values, whereas continuous models need to discretize properly in order to simulate them by using computers.
However, in general, it is not so easy to discretize a continuous dynamical system keeping its mathematical properties properly.
Actually, there exist a lot of discretization of the SIR models which do not inherit mathematical properties of the continuous SIR models, although we kindly hesitate to list examples.
Therefore, from the viewpoint of not mathematics but epidemiology, discrete models of infectious diseases are more preferred.
Meanwhile, proper discretization of dynamical systems has persistently been studied in mathematics especially from the viewpoint of integrable systems, since discrete integrable systems often have more symmetries than continuous ones and hence have mathematically richer structures.
Through the studies, it has gradually been recognized that integrable dynamical systems frequently possess integrable discretization.
Integrable discretization of a continuous integrable system means a discrete dynamical system possessing sufficiently number of conserved quantities, uniqueness of the evolution and convergency to the continuous system with respect to the continuous limit.
Therefore, we can say that integrable discretization properly preserves mathematical properties such as conserved quantities, exact solutions and symmetries of the continuous system.
There exist several recipes to construct integrable discretization from continuous integrable systems \cite{Suris03,Hirota09,Duistermaat10}.
By using the recipes, Willox et al. \cite{WGCR03} and Satsuma et al. \cite{SWRGC04} consecutively proposed integrable discretization of the SIR models in the early 2000s.
They also proposed ultradiscrete analogues, \textit{i.e.}, cellular automata, of the SIR models by applying the procedure of ultradiscretization \cite{TTMJ96}.
Furthermore, Tanaka and Maruno recently proposed another integrable discretization of the original SIR model by applying the hodograph transformation \cite{TM22}.

In this paper, we concentrate on integrable discretization of the SIR model with vaccination firstly proposed in 1970s as an extension of the original SIR model to estimate the effect of vaccination during epidemics \cite{Hethcote74,Dietz75,Schenzle84}.
This model possesses additional two parameters, the birth/death rate and the vaccination rate, to the original SIR model and inherits the total population density as the conserved quantity.
Unfortunately, the SIR model with vaccination does not possess the second conserved quantity in general.
However, when the birth/death rate vanishes the model acquires the second conserved quantity which assures integrability of the model.
We consider this model with vanishing the birth/death rate and propose an integrable discretization which possesses the same conserved quantities as the continuous model by using intersection structure of non-algebraic curves arising from the conserved quantities.
The method of discretization via curve intersection is widely used in integrable systems.
The most famous discrete integrable system derived by applying the method of curve intersection is the celebrated QRT system \cite{QRT89}, which is arising from the intersection of an elliptic curve and a horizontal and a vertical lines as a realization of the additive group structure of an elliptic surface.
It should be noted that the method of curve intersection is even applicable to tropical curves \cite{MS15}, and we can find ultradiscrete integrable systems \cite{IT08,Nobe08,Nobe11,Nobe13}. 
Also remark that tropical curves are algebraic curves since they are defined by ideals of tropical polynomial rings.
Although the method of curve intersection has so far not been widely applied to non-algebraic curves, we diligently apply the method to the intersection of non-algebraic curves relevant to the conserved quantities and find the integrable discretization of the SIR model with vaccination.
The discrete analogue of the SIR model with vaccination is implicitly provided by the intersection of the non-algebraic invariant curve and an another non-algebraic curve; nevertheless the explicit form is proposed by means of the Lambert W function, a multivalued complex function.
Furthermore, single-valuedness of the Lambert W function restricted on the positive real axis assures the uniqueness of the forward/backward evolution of the discrete model. 
Due to the common conserved quantities of the continuous and the discrete SIR models with vaccination, the trajectory of the discrete model forms the same invariant curve as the continuous one.
By using this property, we construct the exact solution to the continuous SIR model with vaccination via the discrete model which parametrizes the common invariant curve.

This paper is organized as follows.
In \S \ref{sec:ISIRM}, we introduce the SIR model with vaccination.
We suppose that the birth/death rate vanishes, and deduce the second conserved quantity provided by a non-algebraic function.
The SIR model with vaccination and without birth/death is abbreviated to the SIRv model.
The integrability of the SIRv model is investigated in terms of Abel's equation of the first kind.
Abel's equation of the first kind thus obtained reduces to the exact differential equation whose potential is explicitly expressed by the second conserved quantity.
In the case where the vaccination rate vanishes, \textit{i.e.}, that of the original SIR model, we obtain Riccati (Bernoulli) equation instead of Abel's equation of the first kind, which is linearizable and hence is integrable.
Remark that linearizability is intrinsic property of the original SIR model and is not inherited to the SIRv model.
In \S \ref{sec:DSIRM}, we investigate the discretization of the SIRv model which preserves integrability, that is, possesses time-reversibility and the conserved quantities.
We show that there exists an integrable discretization of the SIRv model whose conserved quantities are the same as the ones of the continuous model and which is time-reversible as well.
We also find two kinds of integrable discretization of the original SIR model both of which also possess the time-reversibility and the same conserved quantity as the continuous model.
In particular, the time evolution of one of the discrete SIR model is linearized on the plane and is demonstrated by a non-autonomous parallel translation of the line intersecting the invariant curve.
In addition, through the integrable discretization, we construct parametric solutions to the continuous SIR and SIRv models in terms of the Lambert W function.
\S \ref{sec:CONCL} is devoted to concluding remarks.

\section{Integrability of SIR models}
\label{sec:ISIRM}

\subsection{SIR model with vaccination}
\label{subsec:SIRv}

The following system of ODEs governing the dynamics of infectious diseases is called the SIR model with vaccination \cite{Hethcote74,Dietz75,Schenzle84,KNRR21}:
\begin{subequations}
\begin{align}
x^\prime
&=
-\beta xy-\nu x-\mu x+\mu q_1,
\label{eq:SIRvG1}
\\
y^\prime
&=
\beta xy-\gamma y-\mu y,
\label{eq:SIRvG2}
\\
z^\prime
&=
\gamma y+\nu x-\mu z,
\label{eq:SIRvG3}
\end{align}
\end{subequations}
where $x=x(t)$, $y=y(t)$ and $z=z(t)$ are the population densities of the susceptible, the infected and the recovered relevant to the disease, respectively.
We assume that the functions $x$, $y$ and $z$ are real analytic and the ${}^\prime$ denotes the derivative with respect to time $t$.
The non-negative parameters $\beta$, $\gamma$, $\mu$ and $\nu$ respectively stand for the transmission rate, the recovery rate, the birth/death rate and the vaccination rate.
More precisely, the $\mu$ with positive sign stands for the birth rate and that with negative sign for the death rate.
The function $q_1:\R^3\to\R$ in $x,y,z$ is the total population density:
\begin{align*}
 q_1(x,y,z):=x+y+z.
\end{align*}
The equality of the birth and the death rates enables $q_1$ to conserve with respect to $t$; the sum of all equations (\ref{eq:SIRvG1}-\ref{eq:SIRvG3}) leads to
\begin{align*}
q_1^\prime
&=
x^\prime+y^\prime+z^\prime
=
\mu\left(q_1-x-y-z\right)
=
0.
\end{align*}
Thus, the total population density $q_1$ is the first conserved quantity of the dynamical system.

Remark that the population of the death caused by the disease can be included in $z$ being considered as the population density of not the recovered but the removed.
Hence, we can consider that $\mu$ represents the death rate not caused by the disease.
Therefore, it is natural to consider that the birth/death rate $\mu$ is relatively small compared with the other parameters.
Furthermore, in general, the SIR model with vaccination (\ref{eq:SIRvG1}-\ref{eq:SIRvG3}) does not possess the second conserved quantity, whereas the case where the birth/death rate $\mu$ vanishes does; if the effect other than the disease is excluded by assuming $\mu=0$ then the SIR model with vaccination acquires integrability \cite{HLM14}.

\begin{theorem}\label{thm:IntgrbtySIRwV}\normalfont
Assume $\mu=0$.
Then the SIR model with vaccination (\ref{eq:SIRvG1}-\ref{eq:SIRvG3}) has the second conserved quantity $-\beta x-\beta y+\gamma\log x-\nu\log y$.
\end{theorem}

(Proof)\quad
Since $\mu=0$, the SIR model with vaccination (\ref{eq:SIRvG1}-\ref{eq:SIRvG3}) reduces to
\begin{subequations}
\begin{align}
x^\prime
&=
-\beta xy-\nu x,
\label{eq:SIRvO1}
\\
y^\prime
&=
\beta xy-\gamma y,
\label{eq:SIRvO2}
\\
z^\prime
&=
\gamma y+\nu x.
\label{eq:SIRvO3}
\end{align}
\end{subequations}
The total population density $q_1=q_1(x,y,z)$ is still conserved in the system (\ref{eq:SIRvO1}-\ref{eq:SIRvO3}).
Hence, we omit the third equation \eqref{eq:SIRvO3} with assuming the existence of $q_1$.
Also, we see that the parameter $\beta$ can be eliminated from \eqref{eq:SIRvO1} and \eqref{eq:SIRvO2} by applying a proper variable transformation $x\to x/\beta$ and $y\to y/\beta$.
Therefore, we can assume $\beta=1$ without loss of generality.
Thus the three-demensinal system (\ref{eq:SIRvO1}-\ref{eq:SIRvO3}) of ODEs reduces to the two-dimensional system of ODEs
\begin{subequations}
\begin{align}
x^\prime
&=
-xy-\nu x,
\label{eq:SIRv1}
\\
y^\prime
&=
xy-\gamma y,
\label{eq:SIRv2}
\end{align}
\end{subequations}
 on the hyperplane $q_1(x,y,z)=x+y+z$ being constant in $\R^3$

By the sum of \eqref{eq:SIRv1} and \eqref{eq:SIRv2}, we have
\begin{align}
x^\prime
+
y^\prime
&=
-\nu x-\gamma y.
\label{eq:Q2reduce}
\end{align}
If we solve \eqref{eq:SIRv1} and \eqref{eq:SIRv2} for $y$ and $x$, respectively, then we find
\begin{align}
y
&=
-\left(\log x\right)^\prime-\nu
\quad\mbox{and}
\label{eq:yfromSIRv1}
\\
x
&=
\left(\log y\right)^\prime+\gamma.
\label{eq:xfromSIRv2}
\end{align}
Substitute \eqref{eq:yfromSIRv1} and \eqref{eq:xfromSIRv2} into \eqref{eq:Q2reduce}.
Then we get
\begin{align*}
x^\prime
+
y^\prime
&=
-\nu\left(\log y\right)^\prime
+
\gamma\left(\log x\right)^\prime.
\end{align*}
This can be integrated with respect to $t$:
\begin{align*}
-x-y+\gamma\log x-\nu\log y
&=
C,
\end{align*}
where $C$ is the integration constant.
Hence, $q_2:\R^2\to\R$,
\begin{align*}
q_2(x,y):=-x-y+\gamma\log x-\nu\log y
\end{align*}
provides the conserved quantity of (\ref{eq:SIRv1}-\ref{eq:SIRv2}).
If we respectively replace $x$ and $y$ with $\beta x$ and $\beta y$ then (\ref{eq:SIRv1}-\ref{eq:SIRv2}) reduces to (\ref{eq:SIRvO1}-\ref{eq:SIRvO2}).
Therefore, $q_2(\beta x,\beta y)$ gives the second conserved quantity of (\ref{eq:SIRvO1}-\ref{eq:SIRvO3}), \textit{i.e.}, the SIR model with vaccination (\ref{eq:SIRvG1}-\ref{eq:SIRvG3}) imposing $\mu=0$.
\qed

Thus we see that if $\mu$ vanishes then the SIR model with vaccination is integrable since it has sufficient number of conserved quantities $q_1$ and $q_2$.
The system (\ref{eq:SIRv1}-\ref{eq:SIRv2}) of ODEs may be called `the SIR model with vaccination and without birth/death', however, it is simply called the SIRv model, hereafter.
If we substitute $y=-\left(\log x\right)^\prime-\nu$ from \eqref{eq:SIRv1} into \eqref{eq:SIRv2} then the SIRv model reduces to the second order ODE
\begin{align}
\left(\log x\right)^{\prime\prime}
+
\gamma\left(\log x\right)^\prime
-
x^\prime-\nu\left(x-\gamma\right)
=
0.
\label{eq:2ndODE}
\end{align}
We also call the second order ODE \eqref{eq:2ndODE} the SIRv model.

\begin{remark}\label{rem:reductiontoSIR}\normalfont
The system (\ref{eq:SIRvO1}-\ref{eq:SIRvO3}) of ODEs reduces to the original SIR model proposed by Kermack and McKendrick \cite{KM27} if the vaccination rate $\nu$ vanishes:
\begin{subequations}
\begin{align}
x^\prime
&=
-\beta xy,
\label{eq:KMSIR1}
\\
y^\prime
&=
\beta xy-\gamma y,
\label{eq:KMSIR2}
\\
z^\prime
&=
\gamma y.
\label{eq:KMSIR3}
\end{align}
\end{subequations}
Since the both conserved quantities $q_1$ and $q_2$ survive through this reduction, the original SIR model is still integrable.
\end{remark}

\subsection{Invariant curve}
\label{subsec:ICSIRv}

\begin{figure}[htb]
\centering
\includegraphics[scale=1]{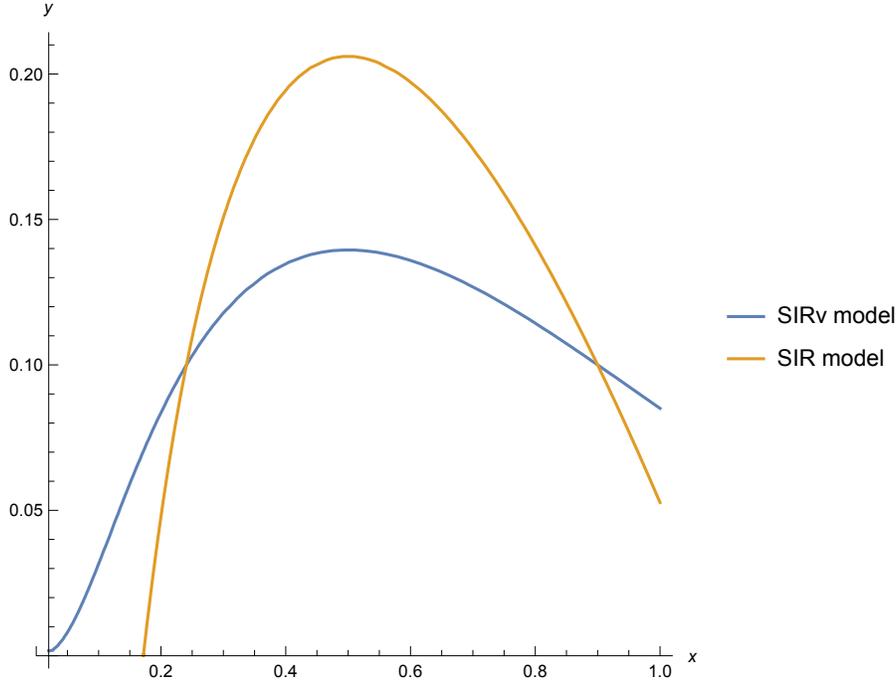}
\caption{The invariant curves $\Gamma$ and $\left.\Gamma\right|_{\nu=0}$ of the SIRv model (\ref{eq:SIRv1}-\ref{eq:SIRv2}) and the SIR model (\ref{eq:KMSIR1}-\ref{eq:KMSIR2}) respectively colored blue and yellow. 
The parameters are chosen as $\beta=1$, $\gamma=0.5$ and $\nu=0.2$, and the initial values as $(x(0),y(0))=(0.9,0.1)$.}
\label{fig:ICSIRv}
\end{figure}

The conserved quantity $q_2$ provides the invariant curve, denoted by $\Gamma$, of the SIRv model (\ref{eq:SIRv1}-\ref{eq:SIRv2}) on the plane $\R^2$ (see figure \ref{fig:ICSIRv}):
\begin{align*}
\Gamma
&:=
\left(
q_2(x,y)-q_2^0
=
-x-y+\gamma\log x-\nu\log y-q_2^0
=
0
\right)
\end{align*}
where $q_2^0=q_2(x(0),y(0))$ is the conserved quantity $q_2$ at the initial point $(x(0),y(0))$.
Remark that the curve $\Gamma$ is a non-algebraic curve since it contains the terms expressed by the logarithmic function.
We see from figure \ref{fig:ICSIRv} that the maximum population density $y$ of the infected is kept lower due to the effect of vaccination.

\begin{figure}[htb]
\centering
\includegraphics[scale=1]{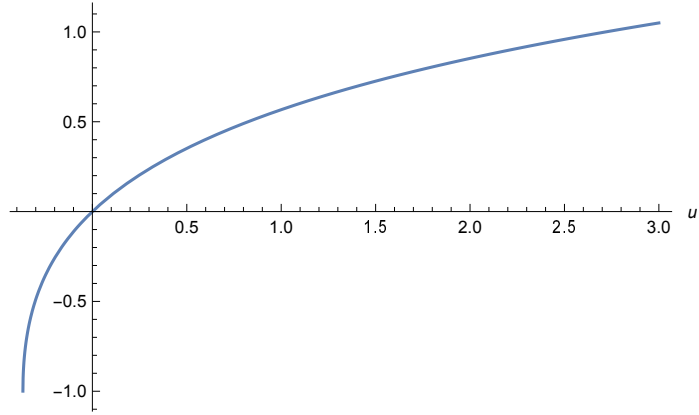}
\caption{The Lambert W function $W(u)$. We choose the branch $W_0(u)\geq-1$ for $-e^{-1}\leq u<0$.}
\label{fig:LambertW}
\end{figure}

Now we show that, although the invariant curve $\Gamma$ is implicit with respect to $y$, we can express it explicitly by means of the Lambert W function \cite{Lambert1758, CGHJK96}.
The Lambert W function, denoted by $W=W(u)$ for $u\in\C$, is defined to be a multivalued complex function satisfying
\begin{align}
W(u)e^{W(u)}=u.
\label{eq:LambertWDef}
\end{align}
For $u\in\R$, $W$(u) reduces to single real-valued on $0\leq u<\infty$, whereas it has two branches $W_0(u)\geq-1$ and $W_1(u)<-1$ on $-e^{-1}\leq u<0$\cite{CGHJK96}.
If we choose the branch $W_0(u)$ then $W(u)$ is a real-valued analytic function defined on $-e^{-1}\leq u<\infty$ (see figure \ref{fig:LambertW}).

We will review several properties of the Lambert W function $W(u)$ necessary for our study.
By definition, we have $W(0)=0$.
We have the derivatives
\begin{align*}
\frac{dW(u)}{du}
&=
\frac{W(u)}{u\left(1+W(u)\right)}
\quad
\mbox{and}
\\
\frac{d^2 W(u)}{du^2}
&=
-\frac{W(u)^2\left(2+W(u)\right)}{u\left(1+W(u)\right)^3}.
\end{align*}
It immediately follows
\begin{align*}
W(u)\leq u
\quad
\mbox{and}
\quad
W(u)=u
\quad
\mbox{if and only if $u=0$.}
\end{align*}
Actually, if we substitute $W(u)=u$ into \eqref{eq:LambertWDef} then we find $u=0$.
Thus the curve $v=W(u)$ intersects the line $v=u$ only at $u=0$.
Moreover, we have
\begin{align*}
\left.\frac{dW(u)}{du}\right|_{u=0}=1
\end{align*}
and
\begin{align*}
\frac{d^2 W(u)}{du^2}
&=
-\frac{W(u)^2\left(2+W(u)\right)}{u\left(1+W(u)\right)^3}
<0
\end{align*}
for $u>0$, which implies
\begin{align*}
\frac{dW(u)}{du}<1
\quad
\mbox{for $u>0$.}
\end{align*}
Similarly, we obtain
\begin{align*}
\frac{dW(u)}{du}>1
\quad
\mbox{for $-e^{-1}\leq u<0$.}
\end{align*}
Thus $W(u)<u$ holds for $-e^{-1}\leq u<\infty$ and $u\neq0$.

The parametrization of $\Gamma$ in terms of $W$ is formulated as follows.
We assume that the vaccination rate $\nu$ is positive.
Then the defining equation $q_2(x,y)-q_2^0=0$ of $\Gamma$ reduces to
\begin{align*}
\frac{y}{\nu}+\log \frac{y}{\nu}
&=
-\frac{x+q_2^0}{\nu}+\frac{\gamma}{\nu}\log x-\log\nu,
\end{align*}
and hence to
\begin{align*}
\frac{y}{\nu}e^{\frac{y}{\nu}}
&=
\frac{1}{\nu}\exp\left(
\frac{\gamma\log x-x-q_2^0}{\nu}
\right).
\end{align*}
We put
\begin{align*}
\zeta(x)
&:=
\frac{1}{\nu}\exp\left(
\frac{\gamma\log x-x-q_2^0}{\nu}
\right).
\end{align*}
Then we have
\begin{align}
y=\nu W(\zeta(x)).
\label{eq:LambertWforIC}
\end{align}
Since $x$ and $y$, the population densities of the susceptible and the infected, are assumed to be positive, any point $(x,y)$ on $\Gamma$ is in the first quadrant.
In addition, $\nu$ is assumed to be positive.
Then we have $\zeta(x)>0$, and hence $W(\zeta(x))$ is positive single-valued as a function in $x$.
Therefore, we find a parametrization of $(x,y)$ on $\Gamma$:
\begin{align*}
(x,y)
&=
\left(x,\nu W(\zeta(x))\right).
\end{align*}

\begin{remark}\label{rem:nu0ornot}\normalfont
Since the limit $\nu\to0$ of $y=\nu W(\zeta(x))$ is singular, we consider the case where $\nu$ vanishes separately.
This is the case of the original SIR model  (\ref{eq:KMSIR1}-\ref{eq:KMSIR3}).
If $\nu=0$ then the conserved quantity $q_2$ reduces to 
\begin{align}
\tilde q_2(x,y)
:=
\left.q_2(x,y)\right|_{\nu=0}
=
-x-y+\gamma\log x.
\label{eq:defofq2}
\end{align}
Thus the invariant curve is provided by
\begin{align*}
\tilde\Gamma
&:=
\left.\Gamma\right|_{\nu=0}
=
\left(
-x-y+\gamma\log x-\tilde q_2^0
=
0
\right),
\end{align*}
where $\tilde q_2^0=\tilde q_2(x(0),y(0))$.
It immediately follows the parametrization of $(x,y)$ on $\tilde\Gamma$:
\begin{align*}
(x,y)
&=
\left(
x,
-x+\gamma\log x-\tilde q_2^0
\right).
\end{align*}
\end{remark}

We will propose a parametric solution to the SIRv model (\ref{eq:SIRv1}-\ref{eq:SIRv2}) in $\S$ \ref{subsec:dSIRv} in terms of the integrable discretization of the SIRv model and the Lambert W function.

\subsection{Abel's equation of the first kind}
\label{subsec:Aeq1}
The SIRv model \eqref{eq:2ndODE} can be related with Abel's equation of the first kind 
\begin{align}
\frac{d\phi}{dx}
&=
f_0
+
f_1\phi
+
f_2\phi^2
+
f_3\phi^3,
\label{eq:AbelODE1k}
\end{align}
where $f_i$ $(i=0,1,2,3)$ is a function in $x$, by applying a proper variable transformation \cite{HLM14}.
We moreover show that the SIRv model is equivalent to an exact differential equation whose potential is relevant to the conserved quantity $q_2$.

Since we assume $\nu\geq0$, we see from \eqref{eq:SIRv1} that the variable $x(t)$ decreases monotonically with respect to $t$.
Thus the inverse function $t=x^{-1}$ of $x(t)$ can be defined.
By introducing a function $\phi:\R\to\R$, we denote the inverse function of $x(t)$ by
\begin{align*}
t&=
\int\phi(x)dx.
\end{align*}

\begin{proposition}\label{prop:Abel}\normalfont
If $x$ solves the SIRv model \eqref{eq:2ndODE} then the function $\phi$ solves Abel's equation of the first kind \eqref{eq:AbelODE1k} equipped with the following $f_i$ ($i=0,1,2,3$)
\begin{align}
f_0(x)
&
=
0,
&
f_1(x)
&=
-\frac{1}{x},
&
f_2(x)
&
=
\gamma-x,
&
f_3(x)
&=
\nu\left(\gamma-x\right)x,
\label{eq:Abelsfs}
\end{align}
and vice versa.
\end{proposition}

(Proof)\quad
First we assume that $x$ solves \eqref{eq:2ndODE}.
By definition of $\phi$, the derivatives of $x$ with respect to $t$ is expressed by using $\phi$ and $d\phi/dx$:
\begin{align}
x^\prime
=
\frac{1}{\phi},
\qquad
x^{\prime\prime}
=
-\frac{1}{\phi^3}\frac{d\phi}{dx}.
\label{eq:philogx}
\end{align}
Substitute \eqref{eq:philogx} into \eqref{eq:2ndODE}.
Then we find
\begin{align*}
\frac{1}{x^2}\left(-\frac{x}{\phi^3}\frac{d\phi}{dx}-\frac{1}{\phi^2}\right)
+
\gamma\frac{1}{x\phi}
-
\frac{1}{\phi}-\nu\left(x-\gamma\right)
&=
0.
\end{align*}
It immediately follows
\begin{align*}
\frac{d\phi}{dx}
&=
-\frac{1}{x}\phi
+
(\gamma-x)\phi^2
-
\nu\left(x-\gamma\right)x\phi^3
\end{align*}
by multiplying $x\phi^3$.
Thus $\phi$ solves \eqref{eq:AbelODE1k} with \eqref{eq:Abelsfs}.

Conversely, if $\phi$ solves \eqref{eq:AbelODE1k}, by substituting $d\phi/dx=-x^{\prime\prime}\phi^3$ into \eqref{eq:AbelODE1k}, we have
\begin{align*}
x^{\prime\prime}\phi^3
+
f_1\phi
+
f_2\phi^2
+
f_3\phi^3
&=
0,
\end{align*}
where we use $f_0=0$.
Division of the above equation by $x\phi^3$ leads to
\begin{align*}
\frac{x^{\prime\prime}}{x}
+
\frac{f_1}{x\phi^2}
+
\frac{f_2}{x\phi}
+
\frac{f_3}{x}
&=
0,
\end{align*}
and hence to
\begin{align*}
\frac{x^{\prime\prime}}{x}
+
\frac{(x^\prime)^2f_1}{x}
+
\frac{x^\prime f_2}{x}
+
\frac{f_3}{x}
&=
0,
\end{align*}
where we use $\phi=1/x^\prime$, the first equation in \eqref{eq:philogx}.
Since $f_i$'s are provided by  \eqref{eq:Abelsfs}, this shows that $x$ solves \eqref{eq:2ndODE}. 
\qed

It is known that Abel's equation of the first kind reduces to the exact differential equation if some conditions are satisfied \cite{MR16}.

\begin{proposition}\label{lem:exactdel}\normalfont
Abel's equation of the first kind \eqref{eq:AbelODE1k} reduces to the exact differential equation if $f_0=0$ and the following condition is satisfied:
\begin{align}
\frac{d}{dx}\frac{f_2}{f_3}
&=
f_1\frac{f_2}{f_3}.
\label{eq:QkP}
\end{align}

\end{proposition}

(Proof)\quad
First, we introduce a new dependent variable $\varphi$:
\begin{align*}
\phi(x)\varphi(x)
&=
\exp
\left(
\int f_1dx
\right).
\end{align*}
By definition of $\varphi$, we have
\begin{align*}
\frac{d\phi}{dx}
&=
\frac{1}{\varphi^2}\left(f_1\varphi-\frac{d\varphi}{dx}\right)\exp\left({\int f_1dx}\right).
\end{align*}
Substitute this into \eqref{eq:AbelODE1k}.
We then find
\begin{align*}
&
\frac{1}{\varphi^2}\left(f_1\varphi-\frac{d\varphi}{dx}\right)
=
\frac{f_1}{\varphi}
+
\frac{f_2}{\varphi^2}\exp\left({\int f_1dx}\right)
+
\frac{f_3}{\varphi^3}\exp\left({2\int f_1dx}\right).
\end{align*}
It follows that, by multiplying $\varphi^3$, we get 
\begin{align*}
\varphi\frac{d\varphi}{dx}
+
f_2\exp\left({\int f_1dx}\right)\varphi
+
f_3\exp\left({2\int f_1dx}\right)
=
0.
\end{align*}
This reduces to
\begin{align}
&
\varphi d\varphi
+
(P+Q\varphi)dx
=
0,
\label{eq:compde}
\end{align}
where we put
\begin{align*}
P(x)
:=
f_3\exp\left({2\int f_1dx}\right)
\quad
\mbox{and}
\quad
Q(x)
:=
f_2\exp\left({\int f_1dx}\right).
\end{align*}

Now we introduce the following integrating factor
\begin{align*}
\varpi(x,\varphi)
&=
\exp\left[-k\left(\varphi+\int Qdx\right)\right],
\end{align*}
where $k$ is a constant.
We show that the equation \eqref{eq:compde} with multiplying the integrating factor $\varpi(x,\varphi)$ is an exact differential equation.
It is necessary to hold
\begin{align}
\frac{\partial}{\partial x}\left(\varphi\varpi\right)
&=
\frac{\partial}{\partial \varphi}\left[(P+Q\varphi)\varpi\right].
\label{eq:exactdecond}
\end{align}
We compute
\begin{align*}
\frac{\partial}{\partial x}\left(\varphi\varpi\right)
&=
-kQ\varphi\varpi
\quad
\mbox{and}
\\
\frac{\partial}{\partial \varphi}\left[(P+Q\varphi)\varpi\right]
&=
(Q-kP)\varpi-kQ\varphi\varpi.
\end{align*}
Thus, if and only if 
\begin{align*}
Q=kP
\end{align*}
the condition \eqref{eq:exactdecond} holds.
This condition reduces to the following equivalent to \eqref{eq:QkP}:
\begin{align*}
\frac{f_2}{f_3}
&=
k\exp\left({\int f_1dx}\right).
\end{align*}

Now assume that \eqref{eq:QkP} is satisfied.
Then the potential $\Psi:\R^2\to\R$ of the exact differential equation equivalent to \eqref{eq:AbelODE1k} is expressed by
\begin{align}
\Psi(x,\varphi)
=
(1+k\varphi)\varpi(x,\varphi).
\label{eq:potential}
\end{align}
Actually, by noticing $Q=kP$, we have
\begin{align}
d\Psi(x,\varphi)
&=
\frac{\partial\Psi}{\partial x}dx
+
\frac{\partial\Psi}{\partial\varphi}d\varphi
=
-k^2\varpi
\left[\varphi d\varphi+(P+Q\varphi)dx
\right].
\label{eq:tautologicalform}
\end{align}
Thus, since $k^2\varpi\not\equiv0$, the exact differential equation $d\Psi=0$ is equivalent to \eqref{eq:compde}, which is nothing but Abel's equation of the first kind \eqref{eq:AbelODE1k}.
\qed

Hereafter, throughout this subsection, we assume $\nu>0$.
This assumption excludes the original SIR model. 
The case where $\nu=0$ will be treated in the next subsection.

\begin{proposition}\label{prop:exacfdeconcrete}\normalfont
Abel's equation of the first kind \eqref{eq:AbelODE1k} equipped with the coefficients \eqref{eq:Abelsfs}, equivalent to the SIRv model  \eqref{eq:2ndODE}, reduces to the exact differential equation whose potential is given by
\begin{align}
\Psi
&=
-\frac{1}{\nu}\exp\left(1-\frac{q_2}{\nu}\right),
\label{eq:potentialcq}
\end{align}
where $q_2$ is the conserved quantity of the SIRv model.
\end{proposition}

(Proof)\quad
First we show that, for the functions $f_i$ ($i=0,1,2,3$) provided by \eqref{eq:Abelsfs}, the condition \eqref{eq:QkP} holds:
\begin{align*}
\frac{d}{dx}\frac{f_2}{f_3}
&=
\frac{d}{dx}\frac{\gamma-x}{\nu\left(\gamma-x\right)x}
=
-\frac{1}{\nu x^2}
=
f_1\frac{f_2}{f_3},
\end{align*}
where we note the assumption $\nu>0$.

Since we have
\begin{align*}
P(x)
&=
f_3\exp\left({2\int f_1dx}\right)
=
\nu\left(\frac{\gamma}{x}-1\right)
\quad\mbox{and}\quad
\\
Q(x)
&=
f_2\exp\left({\int f_1dx}\right)
=
\frac{\gamma}{x}-1,
\end{align*}
we find $k=1/\nu$.
Hence, the integrating factor reduces to
\begin{align*}
\varpi(x,\varphi)
&=
\exp\left(\frac{-\varphi-\gamma\log x+x}{\nu}\right).
\end{align*}

Next we give the potential $\Psi$ by using the conserved quantity $q_2$ of the SIRv model \eqref{eq:2ndODE}.
Remark that we have
\begin{align*}
\varphi
&=
\frac{1}{x\phi}
=
\left(\log x\right)^\prime,
\end{align*}
since $\phi=1/x^\prime$ (see \eqref{eq:philogx}).
For the function $\Psi$ given by \eqref{eq:potential} with $k=1/\nu$, we compute
\begin{align*}
\Psi
&=
\left(1+\frac{1}{\nu}\varphi\right)\varpi(x,\varphi)
\\
&=
-\frac{y}{\nu}\exp\left(\frac{y+\nu-\gamma\log x+x}{\nu}\right),
\end{align*}
where we use $y=-(\log x)^\prime-\nu$ (see \eqref{eq:yfromSIRv1}) reduced from \eqref{eq:SIRv1}.
We then find
\begin{align*}
-\nu\Psi
&=
\exp\left(\frac{x+y-\gamma\log x+\nu\log y+\nu}{\nu}\right)
\\
&=
\exp\left(\frac{\nu-q_2}{\nu}\right).
\end{align*}
This implies \eqref{eq:potentialcq} and completes the proof.
\qed

We confirm that the second order ODE \eqref{eq:2ndODE} is equivalent to the exact differential equation $d\Psi(x,\varphi)=0$.
The derivative of $\varphi$ with respect to $x$ is computed as follows
\begin{align*}
\frac{d\varphi}{dx}
&=
\frac{d}{dx}\frac{1}{x\phi}
=
\frac{-\phi-x\frac{d\phi}{dx}}{x^2\phi^2}
=
-\frac{x^\prime}{x^2}
+\frac{x^{\prime\prime}}{xx^\prime}
=
\frac{1}{x^\prime}\left(\log x\right)^{\prime\prime}.
\end{align*}
Moreover, we have
\begin{align*}
-\frac{\nu^2}{\varpi}\frac{\partial\Psi}{\partial x}
&=
\left(\nu+\left(\log x\right)^\prime\right)\left(\frac{\gamma}{x}-1\right)
\quad\mbox{and}\quad
-\frac{\nu^2}{\varpi}\frac{\partial\Psi}{\partial \varphi}
=
\left(\log x\right)^\prime.
\end{align*}
Therefore, we find
\begin{align*}
-\frac{\nu^2}{\varpi}
\frac{d}{dx}\Psi(x,\varphi(x))
&=
-\frac{\nu^2}{\varpi}\frac{\partial\Psi}{\partial x}
-\frac{\nu^2}{\varpi}\frac{\partial\Psi}{\partial \varphi}\frac{d\varphi}{dx}
\\
&=
\frac{1}{x}
\left[
\left(\nu+\left(\log x\right)^\prime\right)\left(\gamma-x\right)
+
\left(\log x\right)^{\prime\prime}
\right].
\end{align*}
This shows the equivalence of $d\Psi(x,\varphi)=0$ with \eqref{eq:2ndODE}.

\subsection{Riccati equation}
\label{subsec:Riccatieq}
Throughout this subsection, we assume $\nu=0$ and consider the original SIR model as the two-dimensional systems of ODEs
\begin{subequations}
\begin{align}
x^\prime
&=
-xy,
\label{eq:SIR1}
\\
y^\prime
&=
xy-\gamma y,
\label{eq:SIR2}
\end{align}
\end{subequations}
or the second order ODE
\begin{align}
\left(\log x\right)^{\prime\prime}
+
\gamma\left(\log x\right)^\prime
-
x^\prime
=
0.
\label{eq:2ndODESIR}
\end{align}

Then $f_3$ in \eqref{eq:Abelsfs} vanishes and Abel's equation of the first kind \eqref{eq:AbelODE1k} reduces to the Riccati equation
\begin{align}
\frac{d\phi}{dx}
&=
f_0
+
f_1\phi
+
f_2\phi^2
\label{eq:RiccatiODE1k}
\end{align}
whose coefficients are provided by
\begin{align}
f_0(x)
&
=
0,
&
f_1(x)
&=
-\frac{1}{x},
&
f_2(x)
&
=
\gamma-x.
\label{eq:Ruccatifs}
\end{align}
We moreover notice that $f_0=0$, hence \eqref{eq:RiccatiODE1k} is the Bernoulli equation of degree two.

The Bernoulli equation \eqref{eq:RiccatiODE1k} with \eqref{eq:Ruccatifs} reduces to the linear equation in a standard manner.
Introduce the reciprocal $\psi=1/\phi$ of $\phi$.
Then we have
\begin{align*}
\frac{d\psi}{dx}
&=
-\frac{1}{\phi^2}\frac{d\phi}{dx}.
\end{align*}
It follows that \eqref{eq:RiccatiODE1k} reduces to the linear equation
\begin{align}
\frac{d\psi}{dx}-\frac{1}{x}\psi
&=
x-\gamma.
\label{eq:SIRLinear}
\end{align}
The solution to \eqref{eq:SIRLinear} immediately follows
\begin{align}
\psi
&=
\left(x-\gamma\log x+\tilde q_2\right)x,
\label{eq:solSIRmodel}
\end{align}
where $\tilde q_2$ is the conserved quantity of the SIR model defined by \eqref{eq:defofq2} and is the integration constant as well.

The solution \eqref{eq:solSIRmodel} to the Bernoulli euquation \eqref{eq:RiccatiODE1k} with \eqref{eq:Ruccatifs} implicitly provides the exact solution to the SIR model \eqref{eq:2ndODESIR}:
\begin{align*}
t
&=
\int\phi dx
=
\int\frac{1}{\psi}dx
=
\int
\frac{dx}{x\left(x-\gamma\log x+\tilde q_2^0\right)}.
\end{align*}

\section{Discrete SIR models and their integrability}
\label{sec:DSIRM}

First we consider discretization of the original SIR model (\ref{eq:SIR1}-\ref{eq:SIR2}).
Hence, we assume $\nu=0$ from \S\ref{sec:dSIRcq} through \S\ref{sec:GCSIR}.
The case where $\nu>0$ will be investigated in \S\ref{subsec:dSIRv} and \S\ref{sec:ESSIRv}.

\subsection{Integrable discretization of SIR model}
\label{sec:dSIRcq}

We consider the forward differences
\begin{align}
\frac{x^{n+1}-x^n}{\delta}
\qquad
\mbox{and}
\qquad
\frac{y^{n+1}-y^n}{\delta}
\label{eq:forwardD}
\end{align}
with respect to $n$, where $\delta>0$ is the interval of difference and we put
\begin{align*}
x^n
:=
x(t_0+n\delta)
\qquad
\mbox{and}
\qquad
y^n
:=
y(t_0+n\delta)
\end{align*}
for arbitrary $t_0\in\R$.
By applying the Taylor series expansion of $x$ about $t_0+n\delta$, we obtain
\begin{align*}
\frac{x^{n+1}-x^n}{\delta}
&=
\frac{x(t_0+n\delta+\delta)-x(t_0+n\delta)}{\delta}
=
\frac{x^\prime(t_0+n\delta)\delta+o(\delta)}{\delta}
\\
&\to
x^\prime(t_0)
\quad
(\delta\to0)
\end{align*}
and so on.
Since $t_0$ is an arbitrary real number, the forward differences \eqref{eq:forwardD} actually approach the derivatives $x^\prime$ and $y^\prime$ as $\delta\to0$, respectively.

We respectively replace the derivatives $x^\prime$ and $y^\prime$ in (\ref{eq:SIR1}-\ref{eq:SIR2}) with the forward differences in \eqref{eq:forwardD}, and the linear term $\gamma y$ in \eqref{eq:SIR2} with $\gamma y^n$.
Also, we assume that the nonlinear term $xy$ in both \eqref{eq:SIR1} and \eqref{eq:SIR2} is simultaneously discretized into the indefinite $F(x^n,y^n,x^{n+1},y^{n+1};\delta)/\delta$.
Then the system (\ref{eq:SIR1}-\ref{eq:SIR2}) of ODEs is discretized into
\begin{subequations}
\begin{align}
x^{n+1}-x^n
&=
-F(x^n,y^n,x^{n+1},y^{n+1};\delta),
\label{eq:dSIRassump1}
\\
y^{n+1}-y^n
&=
F(x^n,y^n,x^{n+1},y^{n+1};\delta)-\delta\gamma y^n.
\label{eq:dSIRassump2}
\end{align}
\end{subequations}

In order to determine the indefinite nonlinear term $F$, we assume that the system (\ref{eq:dSIRassump1}-\ref{eq:dSIRassump2}) of difference equations has the same conserved quantity $\tilde q_2$ defined by \eqref{eq:defofq2} as the continuous SIR model (\ref{eq:SIR1}-\ref{eq:SIR2}).
Then it is necessary to satisfy
\begin{align*}
0
&
=
\tilde q_2(x^{n+1},y^{n+1})-\tilde q_2(x^n,y^n)
\\
&
=
-\left(x^{n+1}-x^n\right)
-\left(y^{n+1}-y^n\right)
+\gamma\log x^{n+1}-\gamma\log x^n
\\
&
=
\delta\gamma y^n
+\gamma\log x^{n+1}-\gamma\log x^n,
\end{align*}
where the last equality is from the sum of \eqref{eq:dSIRassump1} and \eqref{eq:dSIRassump2}.
This equation can be written as
\begin{align*}
x^{n+1}-x^n
&=
x^n\left(e^{-\delta y^n}-1\right).
\end{align*}
Comparing with \eqref{eq:dSIRassump1}, we uniquely find
\begin{align*}
F(x^n,y^n,x^{n+1},y^{n+1};\delta)
=
-x^n\left(e^{-\delta y^n}-1\right).
\end{align*}
Therefore, in order that the system (\ref{eq:dSIRassump1}-\ref{eq:dSIRassump2}) has the same conserved quantity $\tilde q_2$ as the SIR model, $F$ must be as above.
Hence, \eqref{eq:dSIRassump1} and \eqref{eq:dSIRassump2} respectively reduce to
\begin{subequations}
\begin{align}
x^{n+1}-x^n
&=
x^n\left(e^{-\delta y^n}-1\right),
\label{eq:dSIR1}
\\
y^{n+1}-y^n
&=
-x^n\left(e^{-\delta y^n}-1\right)-\delta\gamma y^n.
\label{eq:dSIR2}
\end{align}
\end{subequations}

Thus we find a candidate for the discrete analogue of the SIR model possessing the conserved quantity $\tilde q_2$.
The following theorem asserts that the system (\ref{eq:dSIR1}-\ref{eq:dSIR2}) actually has several properties to be called an integrable discretization of the SIR model.

\begin{theorem}\label{thm:dSIRthm}\normalfont
The system (\ref{eq:dSIR1}-\ref{eq:dSIR2}) of difference equations 
\begin{enumerate}
\item reduces to the SIR model (\ref{eq:SIR1}-\ref{eq:SIR2}) as $\delta\to0$, and
\item has the conserved quantity $\tilde q_2$.
Moreover, 
\item the forward evolution $(x^n,y^n)\to(x^{n+1},y^{n+1})$ is uniquely determined for any $\delta>0$, and
\item the backward evolution $(x^{n+1},y^{n+1})\to(x^n,y^n)$ is uniquely determined for any $\delta$ such that $0<\delta\leq1/\gamma$. 
\end{enumerate}
\end{theorem}

(Proof)\quad
We have already proven (2), and (3) is clear.
We first show (1).
By applying the Maclaurin series expansion to the exponential function in \eqref{eq:dSIR1}, we have
\begin{align*}
\frac{x^{n+1}-x^n}{\delta}
&=
-x^ny^n-\sum_{\ell=2}^\infty\frac{1}{\ell!}\left(-y^n\right)^\ell\delta^{\ell-1}.
\end{align*}
This implies \eqref{eq:SIR1} as $\delta\to0$.
Also, \eqref{eq:dSIR2} approaches \eqref{eq:SIR2} as $\delta\to0$.

Next consider (4), a necessary condition for integrability.
We solve \eqref{eq:dSIR2} for $y^n$:
\begin{align}
y^n
&=
-\frac{1}{\delta}\log\frac{x^{n+1}}{x^n}.
\label{eq:InvdSIR-I2}
\end{align}
In order to solve (\ref{eq:dSIR1}-\ref{eq:dSIR2}) for $x^n$, we add the equations and substitute \eqref{eq:InvdSIR-I2} into it:
\begin{align*}
x^{n+1}+y^{n+1}
&=
x^n-\Delta\left(\log x^{n+1}-\log x^n\right),
\end{align*}
where we put
\begin{align*}
\Delta
:=
\frac{1-\delta\gamma}{\delta}
=
\frac{1}{\delta}-\gamma.
\end{align*}
Then (\ref{eq:dSIR1}-\ref{eq:dSIR2}) is implicitly solved for $x^n$:
\begin{align}
x^n+\Delta\log x^n
&=
x^{n+1}+y^{n+1}+\Delta\log x^{n+1},
\label{eq:InvdSIR-I1}
\end{align}
The system (\ref{eq:InvdSIR-I2}-\ref{eq:InvdSIR-I1}) of difference equations determines the backward evolution $(x^{n+1},y^{n+1})\to(x^n,y^n)$.

If $\Delta=0$ it is clear that (\ref{eq:InvdSIR-I2}-\ref{eq:InvdSIR-I1}) uniquely determines $(x^n,y^n)$ by $(x^{n+1},y^{n+1})$.
Now we assume $\Delta>0$, that is, $0<\delta<1/\gamma$.
Divide the both side of \eqref{eq:InvdSIR-I1} by $\Delta$, and subtract $\log\Delta$ from them. 
Then we have
\begin{align*}
\frac{x^n}{\Delta}+\log\frac{x^n}{\Delta}
&=
\log\left(\frac{1}{\Delta}e^{\frac{x^{n+1}}{\Delta}}e^{\frac{y^{n+1}}{\Delta}}x^{n+1}\right).
\end{align*}
Therefore, $x^n$ is expressed by using the Lambert W function $W$:
\begin{align*}
x^n=&\Delta W(\theta(x^{n+1},y^{n+1})),
\end{align*}
where we put
\begin{align*}
\theta(x^{n+1},y^{n+1}):=&\frac{1}{\Delta}e^{\frac{x^{n+1}}{\Delta}}e^{\frac{y^{n+1}}{\Delta}}x^{n+1}.
\end{align*}
Since we assume $\Delta>0$, $\theta$ is a positive real-valued function in $x^{n+1}$ and $y^{n+1}$.
Hence, $W$ reduces to single real-valued, and $x^n$ is uniquely determined by $(x^{n+1},y^{n+1})$.

If $\Delta<0$ then $\theta<0$ and hence $W(\theta)$ is double-valued.
Therefore, $x^n$ is not determined uniquely by $(x^{n+1},y^{n+1})$ through \eqref{eq:InvdSIR-I1}.
Thus the backward evolution of (\ref{eq:dSIR1}-\ref{eq:dSIR2}) is uniquely given by (\ref{eq:InvdSIR-I2}-\ref{eq:InvdSIR-I1}) for $0<\delta\leq1/\gamma$.
\qed

\begin{figure}[htb]
\centering
\includegraphics[scale=1]{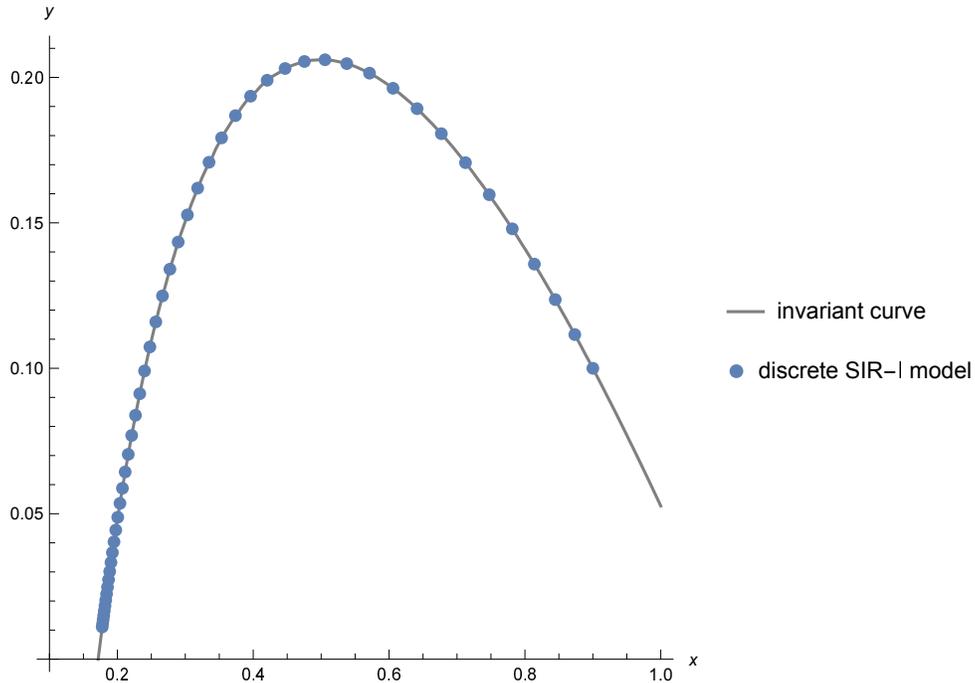}
\caption{The common invariant curve $\tilde\Gamma$ of the continuous (\ref{eq:SIR1}-\ref{eq:SIR2}) and the discrete (\ref{eq:dSIR1}-\ref{eq:dSIR2}) SIR models, and the trajectory of the dSIR-I model (\ref{eq:dSIR1}-\ref{eq:dSIR2}) for $\gamma=0.5$, $(x^0,y^0)=(0.9,0.1)$ and $\delta=0.3$.}
\label{fig:dSIRonICSIR}
\end{figure}

We call the system (\ref{eq:dSIR1}-\ref{eq:dSIR2}) of difference equations the discrete SIR-I model and abbreviate it to dSIR-I model (the dSIR-II model will be obtained later).
Remark that the dSIR-I model is an integrable system of difference equations since it is time-reversible and possesses a sufficient number of conserved quantities.
Figure \ref{fig:dSIRonICSIR} shows a trajectory of dSIR-I model (\ref{eq:dSIR1}-\ref{eq:dSIR2}) on the common invariant curve $\tilde\Gamma$ of the continuous and the discrete SIR models.

By eliminating $y^n$ from \eqref{eq:dSIR1} and \eqref{eq:dSIR2}, we find the second order difference equation
\begin{align*}
\log\frac{x^{n+2}}{x^{n+1}}
-\log\frac{x^{n+1}}{x^n}
&=
\delta(x^{n+1}-x^n)-\gamma\delta\log\frac{x^{n+1}}{x^n}.
\end{align*}
This is a discretization of the second order ODE \eqref{eq:2ndODESIR} and is integrated into
\begin{align*}
\log\frac{x^{n+1}}{x^n}
=
\delta x^n-\gamma\delta\log x^n+\delta\tilde q_2, 
\end{align*}
which is equivalent to the defining equation \eqref{eq:defofq2} of $\tilde q_2$ by noticing \eqref{eq:InvdSIR-I2}.

\subsection{Exact solution to SIR model}
\label{sec:ESSIR}
We propose a parametric solution to the continuous SIR model (\ref{eq:SIR1}-\ref{eq:SIR2}) by using the property that the dSIR-I model (\ref{eq:dSIR1}-\ref{eq:dSIR2}) parametrizes the invariant curve $\tilde\Gamma$.

\begin{corollary}\label{prop:psolSIR}\normalfont
The following $(x(\tau),y(\tau))$ provides the parametric solution to the SIR model (\ref{eq:SIR1}-\ref{eq:SIR2}):
\begin{subequations}
\begin{align}
x(\tau)
&=
x_0e^{-y_0\tau},
\label{eq:psolSIR1}
\\
y(\tau)
&=
x_0+y_0-x_0e^{- y_0\tau}-\gamma y_0\tau,
\label{eq:psolSIR2}
\end{align}
\end{subequations}
where we put $(x_0,y_0):=(x(0),y(0))$ and $\tau=\tau(t)$ is given by
\begin{align}
t
&=
\int
\frac{y_0}{y(\tau)}d\tau.
\label{eq:taut}
\end{align}
\end{corollary}

(Proof)\quad
First remark that the dSIR-I model (\ref{eq:dSIR1}-\ref{eq:dSIR2}) has the conserved quantity $\tilde q_2$ for any $\delta$ which provides the invariant curve $\tilde\Gamma$.
If we put $t^0=0$ then we have $x^1=x(\delta)$ and hence (\ref{eq:dSIR1}) and (\ref{eq:dSIR2}) respectively reduce to
\begin{subequations}
\begin{align}
x^1
&=
x_0e^{-\delta y_0},
\label{eq:psolSIR1p}
\\
y^1
&=
x_0+y_0-x_0e^{-\delta y_0}-\delta\gamma y_0.
\label{eq:psolSIR2p}
\end{align}
\end{subequations}
Manipulating $\delta$, $(x^1,y^1)$ takes any point on the curve $\tilde\Gamma$.
Hence we replace $\delta$ with $\tau(t)$ and find the parametric solution (\ref{eq:psolSIR1}-\ref{eq:psolSIR2}) to the SIR model (\ref{eq:SIR1}-\ref{eq:SIR2}).

Next, we deduce the relation between $\tau=\tau(t)$ and $t$.
By using \eqref{eq:psolSIR1p}  and \eqref{eq:psolSIR2p} with $\delta=\tau$, we obtain
\begin{align*}
\frac{d}{dt}x^1
&=
-x_0y_0e^{-y_0\tau}\tau^\prime
\quad
\mbox{and}
\\
x^1y^1
&=
x_0e^{-y_0\tau}
\left(
-x_0e^{-y_0\tau}+x_0-\gamma y_0\tau+y_0
\right).
\end{align*}
Substituting these into \eqref{eq:SIR1}, we get
\begin{align*}
\tau^\prime
&
=
-\frac{x_0}{y_0}e^{-y_0\tau}+\frac{x_0}{y_0}-\gamma\tau +1
=
\frac{y(\tau)}{y_0},
\end{align*}
which is integrated into \eqref{eq:taut}.
\qed

\subsection{Geometric linearization}
\label{sec:GCSIR}

The time evolution of the dSIR-I model (\ref{eq:dSIR1}-\ref{eq:dSIR2}) is geometrically linearized via a curve intersection.
In terms of the curve intersection, the time evolution of the dSIR-I model is realized as a non-autonomous parallel translation of the line intersecting the invariant curve $\tilde\Gamma$.
The non-autonomous translation means the translation depending on the discrete time $n$. 
Since the linearization of the continuous SIR model via the the Bernoulli equation investigated in \S \ref{subsec:Riccatieq} is non-autonomous (see \eqref{eq:SIRLinear}), it looks natural that the geometric linearization of the dSIR-I model is also non-autonomous. 

\begin{theorem}\label{thm:GeomdSIR}\normalfont
On the $xy$-plane, let the line passing through the point $(x^n,y^n)$ on the invariant curve $\tilde\Gamma$ of the dSIR-I model (\ref{eq:dSIR1}-\ref{eq:dSIR2}) and having slope $-1$ be $\ell$:
\begin{align*}
\ell
&=
\left(
x+y-x^n-y^n=0
\right).
\end{align*}
Apply the variable transformation $(x,y)\to(x,y+\delta\gamma y^n)$ to $\ell$ that deduce the non-autonomous parallel translation of $\ell$; and denote the translated line by $\bar \ell$:
\begin{align*}
\bar \ell
&=
\left(
x+y-x^n-y^n+\delta\gamma y^n=0
\right).
\end{align*}
Then the intersection point of $\tilde\Gamma$ and $\bar \ell$ is $(x^{n+1},y^{n+1})$.
Thus the intersection of $\tilde\Gamma$ and $\bar \ell$ gives the time evolution $(x^n,y^n)\to(x^{n+1},y^{n+1})$ of the dSIR-I model.
\end{theorem}

(Proof)
\quad
Let $x^n:=a$ and $y^n:=b$.
The conserved quantity $\tilde q_2$ provides the invariant curve on the $xy$-plane:
\begin{align*}
\tilde\Gamma
=
\left(
\tilde q_2(x,y)-\tilde q_2(a,b)
=
-x
-y
+\gamma\log x
+a
+b
-\gamma\log a
=
0
\right).
\end{align*}
The defining equation of $\tilde\Gamma$ reduces to
\begin{align*}
\tilde q_2(x,y)-\tilde q_2(a,b)
&=
\gamma\left(\log x-\log a+\delta b\right)
-
\left(x+y-a-b+\delta\gamma b\right).
\end{align*}
Then we see that the intersection point $(x,y)$ of $\tilde\Gamma$ and $\bar \ell$ satisfies
\begin{align*}
\begin{cases}
\log x-\log a+\delta b
=
0,
\\
x+y-a-b+\delta\gamma b
=
0,
\end{cases}
\quad
\Longleftrightarrow
\quad
\begin{cases}
x-a=a(e^{-\delta b}-1),
\\
y-b
=
-a(e^{-\delta b}-1)-\delta\gamma b.
\end{cases}
\end{align*}
This simultaneous system of equations is equivalent to the dSIR-I model (\ref{eq:dSIR1}-\ref{eq:dSIR2}).
Thus the intersection point $(x,y)$ is $(x^{n+1},y^{n+1})$.
\qed

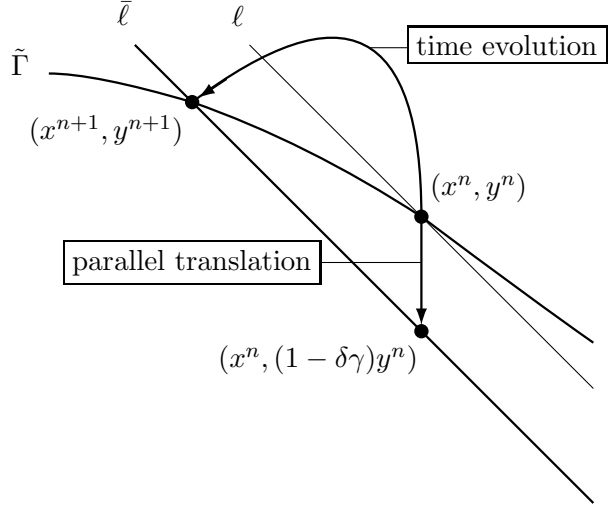
\begin{figure}[htb]
\centering
\unitlength=.03in
\def\arraystretch{1.0}
\begin{picture}(100,100)(0,0)
\put(100,20){\line(-1,1){60}}
\put(52.5,42){\line(1,0){17.5}}
\put(67.6,79.5){\line(-1,0){6.6}}

\linethickness{0.3mm} 
\qbezier(70,50)(70,100)(30,70)
\put(36,74.1){\vector(-3,-2){5}}
\put(70,50){\vector(0,-1){19}}

\linethickness{0.3mm} 
\qbezier(70,50)(50,63)(30,70)
\qbezier(30,70)(13,75)(5,75)
\qbezier(70,50)(90,35)(100,28)

\linethickness{0.3mm} 
\put(100,0){\line(-1,1){80}}
\put(70,50){\circle*{2.5}}
\put(70,30){\circle*{2.5}}
\put(30,70){\circle*{2.5}}
\put(80,55){\makebox(0,0){$(x^n,y^n)$}}
\put(52,25){\makebox(0,0){$(x^n,(1-\delta\gamma)y^n)$}}
\put(15,65){\makebox(0,0){$(x^{n+1},y^{n+1})$}}
\put(38,85){\makebox(0,0){$\ell$}}
\put(18,85.5){\makebox(0,0){$\bar \ell$}}
\put(0,77){\makebox(0,0){$\tilde\Gamma$}}
\put(30,42){\makebox(0,0){\framebox{parallel translation}}}
\put(85,80){\makebox(0,0){\framebox{time evolution}}}
\end{picture}
\caption{
The geometric linearization of the dSIR-I model (\ref{eq:dSIR1}-\ref{eq:dSIR2}). 
The time evolution is realized by the non-autonomous parallel translation of the line $\ell$ to $\bar \ell$ via the intersection with the invariant curve $\tilde\Gamma$.
}
\label{fig:GCdSIR}
\end{figure}

Figure \ref{fig:GCdSIR} shows the geometric linearization of the dSIR-I model.
We see that the non-autonomous parallel translation of the line $\ell$ to $\bar \ell$ generates the time evolution of the dSIR-I model represented by the intersection with the invariant curve $\tilde\Gamma$.
The vertical distance of the translation corresponds to the interval $\delta$ of the time evolution scaled by $\gamma y^n$.
Remark that the backward evolution is also realized as the opposite translation which translates $\bar \ell$ to $\ell$, since the backward evolution is provided by replacing $\delta$ with $-\delta$.
The backward evolution supposed by the geometric realization is uniquely determined for any $\delta>0$.

\subsection{Integrable discretization of SIRv model}
\label{subsec:dSIRv}

Next we consider the case where $\nu>0$, that is, the SIRv model.
Remark that the birth/death rate $\mu$ is still assumed to be $0$.
We will deduce the discretization of the SIRv model (\ref{eq:SIRv1}-\ref{eq:SIRv2}) which possesses the same conserved quantity $q_2$ as (\ref{eq:SIRv1}-\ref{eq:SIRv2}).
Starting from the dSIR-I model (\ref{eq:dSIR1}-\ref{eq:dSIR2}), we consider the following candidate for the discretization of the SIRv model
\begin{align*}
x^{n+1}-x^n
&=
x^n\left(e^{-\delta y^n}-1\right)+G(x^n,y^n,x^{n+1},y^{n+1};\delta),
\\
y^{n+1}-y^n
&=
-x^n\left(e^{-\delta y^n}-1\right)-\delta\gamma y^n,
\end{align*}
where $G$ is to be determined by the assumption that the discretization still possesses the conserved quantity $q_2$.
If we substitute these equations into $q_2(x^{n+1},y^{n+1})-q_2(x^n,y^n)=0$ we find
\begin{align*}
G(x^n,y^n,x^{n+1},y^{n+1};\delta)
&=
\delta\gamma y^n
+\gamma\log\frac{x^{n+1}}{x^n}-\nu\log\frac{y^{n+1}}{y^n}.
\end{align*}
It follows the following system of difference equations as a candidate for the discretization of the SIRv model:
\begin{subequations}
\begin{align}
x^{n+1}-x^n
&=
x^n\left(e^{-\delta y^n}-1\right)+\delta\gamma y^n+\gamma\log\frac{x^{n+1}}{x^n}-\nu\log\frac{y^{n+1}}{y^n},
\label{eq:dSIRvuu1}
\\
y^{n+1}-y^n
&=
-x^n\left(e^{-\delta y^n}-1\right)-\delta\gamma y^n.
\label{eq:dSIRvuu2}
\end{align}
\end{subequations}
By construction, the system (\ref{eq:dSIRvuu1}-\ref{eq:dSIRvuu2}) of difference equations possesses the conserved quantity $q_2$, and it is easy to see that it approaches the SIRv model (\ref{eq:SIRv1}-\ref{eq:SIRv2}) as $\delta\to0$.
However, we can not say that (\ref{eq:dSIRvuu1}-\ref{eq:dSIRvuu2}) is a proper discretization of the SIRv model because the time evolution is not determined uniquely.

Assume $\gamma>0$.
Then $y^{n+1}$ is uniquely determined by $(x^n,y^n)$ in terms of \eqref{eq:dSIRvuu2}.
Also, $x^{n+1}$ is to be determined by $(x^n,y^n)$ and $y^{n+1}$ in terms of \eqref{eq:dSIRvuu1}.  
From \eqref{eq:dSIRvuu1}, we find the explicit form of $x^n$ expressed by the Lambert W function $W$: 
\begin{align*}
x^{n+1}
&=
-\gamma W\left(
-\frac{(y^{n+1})^{\frac{\nu}{\gamma}}}{\gamma}\exp\left(\frac{-x^ne^{-\delta y^n}+q_2^0+x^n+\delta\Delta y^n}{\gamma}\right)
\right).
\end{align*}
Since the argument of $W$ is negative, $W$ is double real-valued.
Hence, $x^{n+1}$ is not determined uniquely (see \S \ref{subsec:ICSIRv}).

In order to find a proper discretization of the SIRv model, we consider the SIRv model (\ref{eq:SIRv1}-\ref{eq:SIRv2}) imposing $\gamma=0$ and $\nu>0$:
\begin{align*}
x^\prime
&=
-xy-\nu x,
\\
y^\prime
&=
xy.
\end{align*}
The discretization of this system of ODEs which possesses the conserved quantity $\left.q_2\right|_{\gamma=0}$ is similarly obtained as (\ref{eq:dSIR1}-\ref{eq:dSIR2}):
\begin{subequations}
\begin{align}
x^{n+1}-x^n
&=
-y^n\left(e^{\delta x^n}-1\right)-\delta\nu x^n,
\label{eq:dSIRvgn1}
\\
y^{n+1}-y^n
&=
y^n\left(e^{\delta x^n}-1\right).
\label{eq:dSIRvgn2}
\end{align}
\end{subequations}

We extend the system (\ref{eq:dSIRvgn1}-\ref{eq:dSIRvgn2}) of difference equations to the case where $\gamma>0$ as in the manner deriving (\ref{eq:dSIRvuu1}-\ref{eq:dSIRvuu2}).
We consider the following candidate for the discretization of the SIRv model imposing $\gamma>0$ and $\nu>0$:
\begin{align*}
x^{n+1}-x^n
&=
-y^n\left(e^{\delta x^n}-1\right)-\delta\nu x^n,
\\
y^{n+1}-y^n
&=
y^n\left(e^{\delta x^n}-1\right)+H(x^n,y^n,x^{n+1},y^{n+1};\delta),
\end{align*}
where $H$ is the indefinite term to be determined.
If we substitute these equations into $q_2(x^{n+1},y^{n+1})-q_2(x^n,y^n)=0$ then $H$ is uniquely determined
\begin{align*}
H(x^n,y^n,x^{n+1},y^{n+1};\delta)
&=
\delta\nu x^n+\gamma\log\frac{x^{n+1}}{x^n}-\nu\log\frac{y^{n+1}}{y^n}.
\end{align*}
It follows the system of difference equations as the candidate for the discretization of the SIRv model
\begin{subequations}
\begin{align}
x^{n+1}-x^n
&=
-y^n\left(e^{\delta x^n}-1\right)-\delta\nu x^n,
\label{eq:dSIRv1}
\\
y^{n+1}-y^n
&=
y^n\left(e^{\delta x^n}-1\right)+\delta\nu x^n+\gamma\log\frac{x^{n+1}}{x^n}-\nu\log\frac{y^{n+1}}{y^n}.
\label{eq:dSIRv2}
\end{align}
\end{subequations}
The equation \eqref{eq:dSIRv2} is explicitly solved for $y^{n+1}$ by using the Lambert W function $W$:
\begin{align}
y^{n+1}
&=
\nu W(\zeta(x^{n+1})),
\label{eq:dSIRvexplicit2}
\end{align}
where
\begin{align*}
\zeta(x^{n+1})
&=
\frac{1}{\nu}\exp\left(\frac{\gamma\log x^{n+1}-x^{n+1}-q_2^0}{\nu}\right)
\end{align*}
and $q_2^0=q_2(x^0,y^0)$.

We then find the following lemma asserting the uniqueness of the time evolution.

\begin{lemma}\normalfont\label{lem:uniquenessFEdSIRv}
In the system (\ref{eq:dSIRv1}-\ref{eq:dSIRv2}) of difference equations, $(x^{n+1}, y^{n+1})$ is uniquely determined by $(x^n,y^n)$ for any $\delta>0$.
\end{lemma}

(Proof)\quad
It is clear that $x^{n+1}$ is uniquely determined by $(x^n,y^n)$ in terms of \eqref{eq:dSIRv1}.
The uniqueness of \eqref{eq:dSIRvexplicit2} is also clear since the Lambert W function $W(\zeta(x^{n+1}))$ is single-valued for $\zeta(x^{n+1})>0$ and $\zeta(x^{n+1})$ is actually positive for $\gamma>0$ and $\nu>0$.
Noticing that \eqref{eq:dSIRvexplicit2} gives the explicit form of \eqref{eq:dSIRv2} completes the proof.
\qed

Thus we see that the forward evolution of (\ref{eq:dSIRv1}-\ref{eq:dSIRv2}) is uniquely determined for any $\delta>0$.
Meanwhile, the uniqueness of the backward evolution is restricted; it is uniquely determined only for sufficiently small $\delta>0$.

\begin{lemma}\normalfont\label{lem:uniquenessBEdSIRv}
In the system (\ref{eq:dSIRv1}-\ref{eq:dSIRv2}) of difference equations, $(x^n,y^n)$ is uniquely determined by $(x^{n+1}, y^{n+1})$ for sufficiently small $\delta>0$.
\end{lemma}

(Proof)\quad
Put $x:=x^n$, $y:=y^n$, $a:=x^{n+1}$ and $b:=y^{n+1}$.
Given $a$, the equation \eqref{eq:dSIRv1} defines a curve, denoted by $\Theta$, on the $xy$-plane:
\begin{align}
\Theta
&:=
\left(
y\left(e^{\delta x}-1\right)-\tilde\Delta x+a=0
\right),
\label{eq:curveTheta}
\end{align}
where we put $\tilde\Delta:=1-\delta\nu$.
Also, given $a$ and $b$, the sum of \eqref{eq:dSIRv1} and \eqref{eq:dSIRv2} reduces to
\begin{align}
-x-y+\gamma\log x-\nu\log y+a+b-\gamma\log a+\nu\log b=q_2(x,y)-q_2(a,b)=0,
\label{eq:iesolve2}
\end{align}
which provides the invariant curve $\Gamma$ on the $xy$-plane.
Hence, the intersection point of the curves $\Theta$ and $\Gamma$ determines $x$ and $y$.
Therefore, we have only to show that there exists exactly an intersection point for $\delta>0$ small enough.

First we consider the curve $\Theta$ provided by \eqref{eq:curveTheta}.
We solve $\Theta$ for $y$:
\begin{align*}
y
&=
\frac{\tilde\Delta x-a}{e^{\delta x}-1}=:g_1(x),
\end{align*}
where we denote the right-hand-side by $g_1(x)$.
Since $\delta>0$ and $x>0$, $y>0$ is equivalent to $x>a/\tilde\Delta$.
Therefore, we consider the smooth function $g_1(x)$ on the interval $(a/\tilde\Delta,\infty)$.
Moreover, we assume $0<\tilde\Delta<1$, that is, $0<\delta<1/\nu$.

The critical point of $g_1(x)$ is obtained by solving
\begin{align}
\frac{dg_1(x)}{dx}
&=
\frac{\left((1-\delta x)\tilde\Delta+\delta a\right)e^{\delta x}-\tilde\Delta}{(e^{\delta x}-1)^2}
=
0,
\label{eq:eqcpcie}
\end{align}
which reduces to
\begin{align*}
\left(\delta x-1-\frac{\delta a}{\tilde\Delta}\right)e^{\delta x-1-\frac{\delta a}{\tilde\Delta}}
&=
-e^{-1-\frac{\delta a}{\tilde\Delta}}.
\end{align*}
Since $-1-\delta a/\tilde\Delta<-1$, we have $-e^{-1}<-e^{-1-\frac{\delta a}{\tilde\Delta}}<0$.
Hence \eqref{eq:eqcpcie} has a unique solution $x=x_1$ expressed by using the real branch $W_0\geq-1$ of the Lambert W function $W$:
\begin{align*}
x_1
&=
\frac{1}{\delta}\left(W_0\left(-e^{-1-\frac{\delta a}{\tilde\Delta}}\right)+1\right)+\frac{a}{\tilde\Delta}.
\end{align*}
Then we see that $x_1-a/\tilde\Delta>0$ holds.
This implies that there exists exactly a critical point $x_1$ of $g_1(x)$ on the interval $(a/\tilde\Delta,\infty)$.

By computing the second derivative of $g_1(x)$, we find
\begin{align*}
\left.\frac{d^2g_1(x)}{dx^2}\right|_{x=x_1}
&=
-\delta\tilde\Delta
\left(
\frac{1}{W_0\left(-e^{-1-\frac{\delta a}{\tilde\Delta}}\right)+1}-1
\right)
<0,
\end{align*}
where we use $-1<W_0\left(-e^{-1-\frac{\delta a}{\tilde\Delta}}\right)<0$.
Therefore, $g_1(x)$ has a unique local maximum at the critical point $x_1$.

The local maximum $y_1=g_1(x_1)$ is also expressed by the branch $W_0$ as follows.
Since $x=x_1$ satisfies \eqref{eq:eqcpcie}, we have
\begin{align*}
e^{\delta x_1}-1
=
\frac{\delta\left(\tilde\Delta x_1-a\right)}{1-\delta\left(\tilde\Delta x_1-a\right)}.
\end{align*}
Then we find
\begin{align}
y_1
&=
\frac{\tilde\Delta x_1-a}{e^{\delta x_1}-1}
=
-\frac{\tilde\Delta}{\delta} W_0\left(-e^{-1-\frac{\delta a}{\tilde\Delta}}\right).
\label{eq:localmaximumy1}
\end{align}

By \eqref{eq:localmaximumy1}, as a function in $\delta$, $y_1$ is monotonically decreasing and satisfies
\begin{align*}
\lim_{\delta\to0}y_1&=\infty
\quad
\mbox{and}
\quad
\lim_{\delta\to1/\nu}y_1=0,
\end{align*}
since we have
\begin{align*}
\lim_{\delta\to0}\tilde\Delta
&=
1
\quad
\mbox{and}
\quad
\lim_{\delta\to1/\nu}\tilde\Delta
=
0
\end{align*}
and hence have
\begin{align*}
\lim_{\delta\to0}W_0\left(-e^{-1-\frac{\delta a}{\tilde\Delta}}\right)
&=
W_0(-e^{-1})
=
-1
\quad
\mbox{and}
\quad
\lim_{\delta\to1/\nu}W_0\left(-e^{-1-\frac{\delta a}{\tilde\Delta}}\right)
=
W(0)
=
0.
\end{align*}
It immediately follows that the curve $\Theta$ approaches the vertical line $x=a$ as $\delta\to0$, which is consistent with the convergence of $x$ to $a$:
\begin{align*}
\lim_{\delta\to0}\left|x^n-x^{n+1}\right|
&=
\lim_{\delta\to0}\left|x-a\right|
=
0.
\end{align*}
The curve $\Theta$ similarly approaches the horizontal line $y=0$ as $\delta\to1/\nu$.

We next investigate the invariant curve $\Gamma$.
The equation \eqref{eq:iesolve2} is solved for $y$:
\begin{align*}
y
&=
\nu
W\left(
\zeta(x)
\right)=:g_2(x),
\end{align*}
where we denote the right-hand-side by $g_2(x)$.
Remark that, since $\nu>0$ and hence $\zeta(x)>0$, $g_2(x)$ is a smooth function defined on $x>0$.

If we solve the equation
\begin{align*}
\frac{dg_2(x)}{dx}
&=
\nu \frac{d W(\zeta)}{d\zeta}\frac{d\zeta(x)}{dx}
=
\frac{W(\zeta)}{1+W(\zeta)}\left(\frac{\gamma}{x}-1\right)
=
0,
\end{align*}
we obtain the unique critical point $x_2=\gamma$ of $g_2(x)$, where we use the fact that $W(\zeta)>0$ for any $\zeta(x)>0$ since $\zeta(x)>0$ for any $x>0$.
We also find
\begin{align*}
\left.\frac{d^2g_2(x)}{dx^2}\right|_{x=x_2}
&=
-\frac{W(\zeta(x_2))}{\gamma(1+W(\zeta(x_2)))}
<0.
\end{align*}
Therefore, $g_2(x)$ has a unique local maximum at the critical point $x_2=\gamma$.
The local maximum $y_2=g_2(x_2)$ is given by
\begin{align*}
y_2
&=
\nu W(\zeta(x_2))
=
\nu W\left(
\frac{b}{\nu}\left(\frac{\gamma}{a}\right)^{\frac{\gamma}{\nu}}e^{\frac{a+b-\gamma}{\nu}}
\right).
\end{align*}

\begin{figure}[htb]
\centering
\subfigure[$\delta=0.75$]{
\includegraphics[scale=.7]{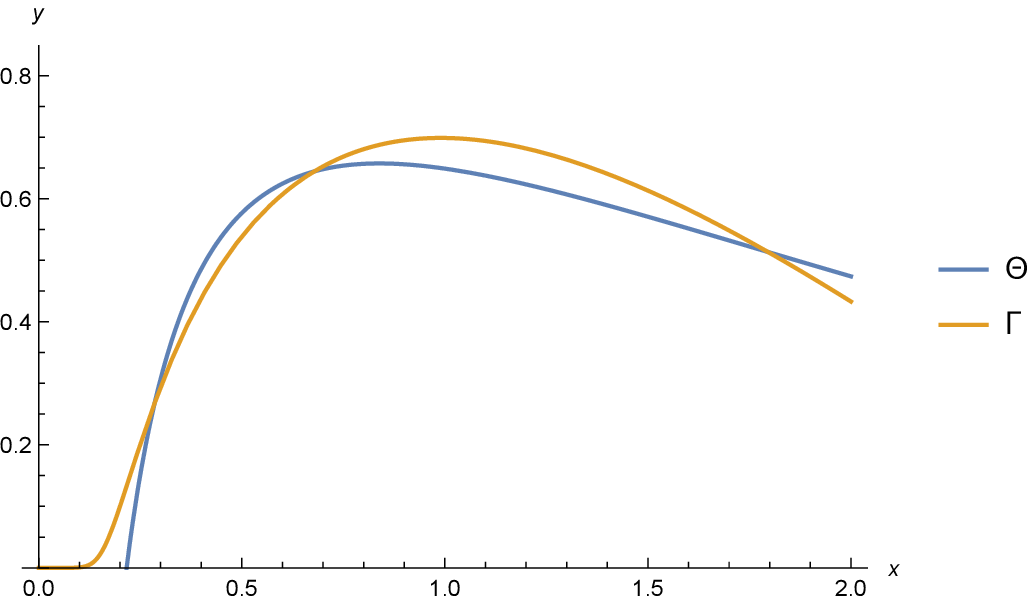}}
\quad 
\subfigure[$\delta=0.65$]{
\includegraphics[scale=.7]{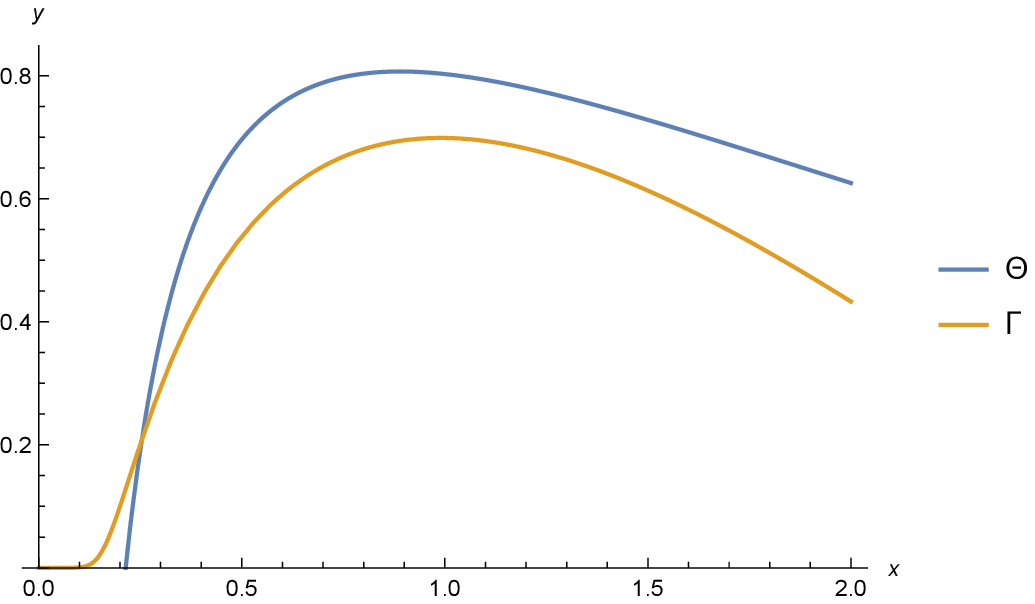}}
\caption{The curves $\Theta$ provided by \eqref{eq:curveTheta} and $\Gamma$ by \eqref{eq:iesolve2} for $\gamma=0.99$, $\nu=0.1$ and $(a,b)=(0.2,0.1)$.
In (a), $\Theta$ and $\Gamma$ intersect at three points for $\delta=0.75$, whereas at a point for $\delta=0.65$ in (b).}
\label{fig:fgintersect3points}
\end{figure}

Now we enumerate the intersection points of the curves $\Theta=(y-g_1(x)=0)$ and $\Gamma=(y-g_2(x)=0)$.
Consider the asymptotic behaviors of the functions $g_1(x)$ and $g_2(x)$ as $x$ approaches infinity. 
For $g_1(x)$, we have
\begin{align*}
g_1(x)
&=
\frac{\tilde\Delta x-a}{e^{\delta x}-1}
>
\left(\tilde\Delta x-a\right)e^{-\delta x}.
\end{align*}
For $g_2(x)$, we easily see that it satisfies
\begin{align*}
g_2(x)
&=
\nu W\left(
\frac{1}{\nu}
x^{\frac{\gamma}{\nu}}
e^{-\frac{x+q_2}{\nu}}
\right)
<
x^{\frac{\gamma}{\nu}}
e^{-\frac{x+q_2}{\nu}},
\end{align*}
where we use the fact $W(u)<u$ for $u>0$ (see \S \ref{subsec:ICSIRv}).
Noticing $0<\delta<1/\nu$, we see that the following holds for sufficiently large $x$
\begin{align}
g_1(x)
&>
\left(\tilde\Delta x-a\right)e^{-\delta x}
>
x^{\frac{\gamma}{\nu}}
e^{-\frac{x+q_2}{\nu}}
>
g_2(x).
\label{eq:g1g2inequality}
\end{align}
Meanwhile, at the other end $x=a/\tilde\Delta$ of the interval $(a/\tilde\Delta,\infty)$ on which both $g_1(x)$ and $g_2(x)$ are positive, we have the opposite inequality between $g_1(x)$ and $g_2(x)$:
\begin{align*}
g_1\left(\frac{a}{\tilde\Delta}\right)
&=
0
<
\nu W\left(\frac{a}{\tilde\Delta}\right)
=
g_2\left(\frac{a}{\tilde\Delta}\right).
\end{align*}
Therefore, since both $g_1(x)$ and $g_2(x)$ are continuous functions, there exists at least an intersection point of the curves $\Theta$ and $\Gamma$ on the interval $(a/\tilde\Delta,\infty)$.
Remark that, for $\delta$ not small enough, there may exist plural intersection points (see figure \ref{fig:fgintersect3points}).

If $\delta\to0$ then the curve $\Theta$ goes upward keeping $g_1(a/\tilde\Delta)=0$.
The local maximum $y_1=g_1(x_1)$ (see \eqref{eq:localmaximumy1}) increases as large as possible, while the invariant curve $\Gamma$ is fixed.
Furthermore, we see from \eqref{eq:eqcpcie} that the slope of the tangent line to $\Theta$ at the left end $x=a/\tilde\Delta$ of the interval $(a/\tilde\Delta,\infty)$,
\begin{align*}
\left.\frac{dg_1(x)}{dx}\right|_{x=a/\tilde\Delta}
&=
\frac{\tilde\Delta}{e^{\delta x}-1},
\end{align*}
increases as large as possible as $\delta\to0$. 
Thus, for sufficiently small $\delta>0$, $\Theta$ must intersect $\Gamma$ only at a point on the interval $(a/\tilde\Delta,x_1)$ (see figure \ref{fig:fgintersect3points}).

Suppose $x>x_1$ and fix it for a while.
Since $g_2(x)$ is smooth, there exists a positive real number $M(x)$ such that $g_2(x)<M(x)$.
For $M(x)$, there exists $\delta=\delta(x)\in(0,1/\nu)$ such that $g_1(x)>M(x)$, since $g_1(x)$ increases as large as possible as $\delta\to0$.
Moreover, since $g_2(x)$ is independent of $\delta$, for any $x>x_1$, there exits $\delta=\delta(x)$ satisfying $g_1(x)>M(x)>g_2(x)$. 
Now we move $x$ and let $\delta_0:=\inf_{x_1<x}\delta(x)$.
Since $g_1(x)>g_2(x)$ holds for sufficiently large $x$ (see \eqref{eq:g1g2inequality}), the infimum $\delta_0$ is attained at a finite $x$. 
Therefore, $\delta_0\neq0$ and $\delta_0>0$ holds.
Thus, for any $0<\delta<\delta_0$, we have $g_1(x)>g_2(x)$ for any $x>x_1$.
Hence, $\Theta$ does not intersect $\Gamma$ on the interval $(x_1,\infty)$ (see figure \ref{fig:fgintersect3points}).
Therefore, there exists exactly an intersection point of the curves $\Theta$ and $\Gamma$ for sufficiently small $\delta>0$.
It follows that the backward evolution is uniquely determined for such $\delta$.
\qed

Thus we see from above lemmas that the system (\ref{eq:dSIRv1}-\ref{eq:dSIRv2}) of difference equations is time-reversible for sufficiently small $\delta>0$.
We moreover obtain the following theorem asserting the integrability of the system (\ref{eq:dSIRv1}-\ref{eq:dSIRv2}) of difference equations.

\begin{theorem}\normalfont\label{thm:dSIRvUTE}
The system (\ref{eq:dSIRv1}-\ref{eq:dSIRv2}) of difference equations 
\begin{enumerate}
\item reduces to the SIRv model (\ref{eq:SIRv1}-\ref{eq:SIRv2}) as $\delta\to0$, and
\item has the conserved quantity $q_2$.
Moreover, 
\item the forward evolution $(x^n,y^n)\to(x^{n+1},y^{n+1})$ is uniquely determined for any $\delta>0$, and
\item the backward evolution $(x^{n+1},y^{n+1})\to(x^n,y^n)$ is uniquely determined for sufficiently small $\delta>0$. 
\end{enumerate}
\end{theorem}

(Proof)\quad
(1)\
The property that \eqref{eq:dSIRv1} approaches \eqref{eq:SIRv1} as $\delta\to0$ is similarly shown to theorem \ref{thm:dSIRthm} (1).
By applying the Taylor series expansion, \eqref{eq:dSIRv2} reduces to
\begin{align*}
y^\prime(t_n)\delta+o(\delta)
&=
y(t_n)\left(\delta x(t_n)+o(\delta)\right)+\delta\nu x(t_n)\\
&\qquad
+\gamma\log_1\left(1+\frac{x^\prime(t_n)}{x(t_n)}\delta+o(\delta)\right)
-\nu\log_1\left(1+\frac{y^\prime(t_n)}{y(t_n)}\delta+o(\delta)\right)
\\
&=
\delta x(t_n)y(t_n)+\delta\nu x(t_n)
+\gamma\frac{x^\prime(t_n)}{x(t_n)}\delta
-\nu\frac{y^\prime(n)}{y(t_n)}\delta+o(\delta).
\end{align*}
It follows that we have
\begin{align*}
y^\prime(t_n)
&=
x(t_n)y(t_n)+\nu x(t_n)
+\gamma\left(\log x(t_n)\right)^\prime
-\nu\left(\log y(t_n)\right)^\prime+o(1),
\end{align*}
which approaches
\begin{align*}
y^\prime
&=
xy+\nu x
+\gamma\left(\log x\right)^\prime
-\nu\left(\log y\right)^\prime
\end{align*}
as  $\delta\to0$.
Now we substitute $(\log x)^\prime=-y-\nu$, which is reduced from \eqref{eq:SIRv1}, into this equation.
We then find
\begin{align*}
(y+\nu)\left(y^\prime-xy+\gamma y\right)
&=
0.
\end{align*}
Since $y\not\equiv-\nu$, we obtain
\begin{align*}
y^\prime-xy+\gamma y
=
0.
\end{align*}
This is nothing but \eqref{eq:SIRv2}.

(2)\
It has already been proven in lemma \ref{lem:uniquenessBEdSIRv} (see \eqref{eq:iesolve2}).

(3)\
It is clear from lemma \ref{lem:uniquenessFEdSIRv}.

(4)\
It is also clear from lemma \ref{lem:uniquenessBEdSIRv}, which completes the proof.
\qed

\begin{figure}[htb]
\centering
\includegraphics[scale=1]{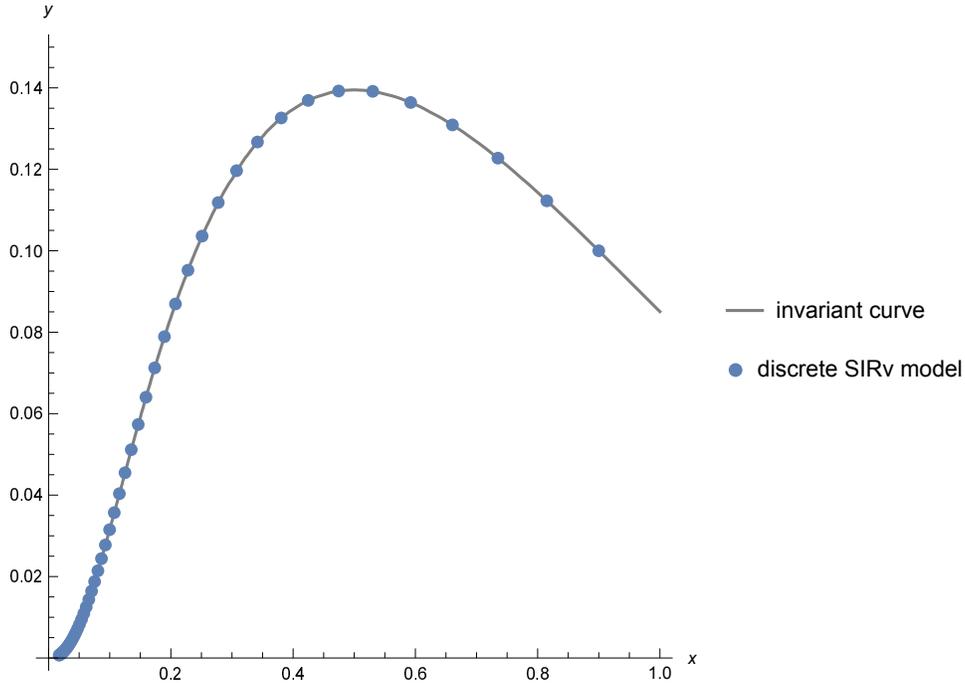}
\caption{The trajectory of the dSIRv model (\ref{eq:dSIRv1}-\ref{eq:dSIRv2}) and the invariant curve $\Gamma$ for $\gamma=0.5$, $\nu=0.2$, $(x^0,y^0)=(0.9,0.1)$ and $\delta=0.3$.}
\label{fig:dSIRv}
\end{figure}

Thus we can say that the system (\ref{eq:dSIRv1}-\ref{eq:dSIRv2}) of difference equations is an integrable discrete analogue of the SIRv model.
Therefore, we call it the discrete SIRv model and abbreviate it to the dSIRv model.
Figure \ref{fig:dSIRv} shows an example of the time evolution of the dSIRv model in which the trajectory is on the common invariant curve $\Gamma$ of the continuous and the discrete SIRv models.

\begin{figure}[htb]
\centering
\includegraphics[scale=1]{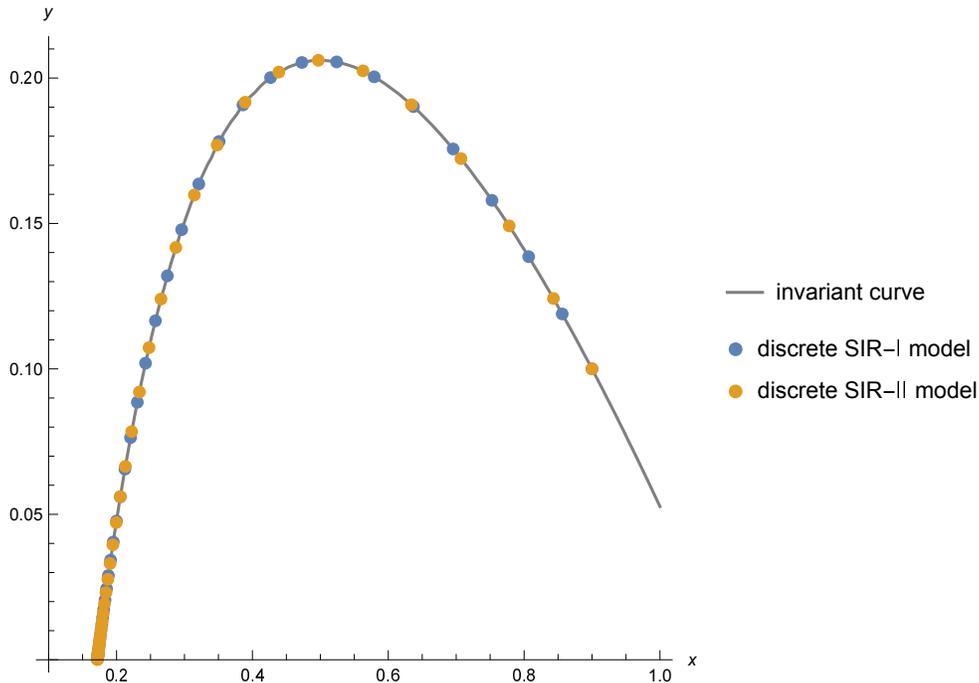}
\caption{The trajectories of the dSIR-I (\ref{eq:dSIR1}-\ref{eq:dSIR2}) and the dSIR-II (\ref{eq:dSIRII1}-\ref{eq:dSIRII2}) models, respectively denoted by the blue and the yellow points,  for $\gamma=0.5$, $(x^0,y^0)=(0.9,0.1)$ and $\delta=0.5$. Both trajectories are on the common invariant curve $\tilde\Gamma$.}
\label{fig:dSIRIIIonICSIR}
\end{figure}

Now suppose $\nu=0$. 
Then the system (\ref{eq:dSIRv1}-\ref{eq:dSIRv2}) of difference equations reduces to
\begin{subequations}
\begin{align}
x^{n+1}-x^n
&=
-y^n\left(e^{\delta x^n}-1\right),
\label{eq:dSIRII1}
\\
y^{n+1}-y^n
&=
y^n\left(e^{\delta x^n}-1\right)+\gamma\log\frac{x^{n+1}}{x^n}.
\label{eq:dSIRII2}
\end{align}
\end{subequations}
This provides a discretization of the original SIR model different from the dSIR-I model (\ref{eq:dSIR1}-\ref{eq:dSIR2}).
Therefore, we call the system (\ref{eq:dSIRII1}-\ref{eq:dSIRII2}) of difference equations the dSIR-II model.
Figure \ref{fig:dSIRIIIonICSIR} shows the trajectories of the dSIR-I and -II models  starting from the same initial point, the right-most point, and on the unique invariant curve $\tilde\Gamma$.
We see that the two discretization provide the different time evolutions of the same initial point.

\subsection{Exact solution to SIRv model}
\label{sec:ESSIRv}

We propose a parametric solution to the SIRv model by using the dSIRv model which parametrizes the invariant curve $\Gamma$.

\begin{corollary}\label{prop:psolSIRv}\normalfont
Assime $\nu>0$.
Then the following $(x(\tau),y(\tau))$ provides the parametric solution to the SIRv model (\ref{eq:SIRv1}-\ref{eq:SIRv2})
\begin{align}
x(\tau)
&=
x_0+y_0-\nu x_0\tau
-y_0e^{x_0\tau},
\label{eq:psolSIRv1}
\\
y(\tau)
&=
\nu W(\zeta),
\label{eq:psolSIRv2}
\end{align}
where we put $(x_0,y_0)=(x(0),y(0))$, $q_2^0=q_2(x_0,y_0)$ and
\begin{align*}
\zeta
=&
\frac{1}{\nu}\exp\left(
\frac{\gamma\log x(\tau)-x(\tau)-q_2^0}{\nu}
\right).
\end{align*}
The parameter $\tau=\tau(t)$ is given by
\begin{align*}
t
&=
\int\frac{x_0\left(\nu+y_0e^{x_0\tau}\right)}{x(\tau)(\nu+y(\tau))}d\tau.
\end{align*}
\end{corollary}

(Proof)\quad
As is the case with the dSIR-I model, the dSIRv model (\ref{eq:dSIRv1}-\ref{eq:dSIRv2}) with $n=0$ reduces to  \eqref{eq:psolSIRv1} and
\begin{align}
y
&=
\nu x_0\tau
+y_0e^{x_0\tau}+\gamma\log\frac{x}{x_0}-\nu\log\frac{y}{y_0}
\label{eq:psolSIRv2o}
\end{align}
by imposing $(x,y):=(x^1,y^1)$, $(x_0,y_0):=(x^0,y^0)$ and $\tau:=\delta$.
Then $(x,y)$ takes any point on the invariant curve $\Gamma$ by manipulating $\tau$.
In order to relate $\tau$ with $t$ explicitly, we solve \eqref{eq:psolSIRv2o} for $y$ as \eqref{eq:psolSIRv2} by using the Lambert W function (see \eqref{eq:LambertWforIC}).
Thus, $y$ is uniquely determined by $(x_0,y_0)$ on the first quadrant since $\zeta>0$ for $\nu>0$.
Hence, (\ref{eq:psolSIRv1}-\ref{eq:psolSIRv2}) provides a parametric solution to SIRv model (\ref{eq:SIRv1}-\ref{eq:SIRv2}).

By differentiating $x(\tau)$ with respect to $t$, we have
\begin{align*}
x^\prime
&=
\frac{dx(\tau)}{d\tau}\tau^\prime
=
-x_0\left(\nu+y_0e^{x_0\tau}\right)\tau^\prime
\end{align*}
Substitute $x^\prime$ into \eqref{eq:SIRv1}.
We then find
\begin{align*}
-x_0\left(\nu+y_0e^{x_0\tau}\right)\tau^\prime
&=
-x(\tau)(\nu+y(\tau)).
\end{align*}
It immediately follows
\begin{align*}
t
&=
\int\frac{x_0\left(\nu+y_0e^{x_0\tau}\right)}{x(\tau)(\nu+y(\tau))}d\tau, 
\end{align*}
which completes the proof.
\qed

\section{Concluding remarks}
\label{sec:CONCL}

We investigate the integrable discretization of the SIR model with vaccination (\ref{eq:SIRvG1}-\ref{eq:SIRvG3}) imposing the condition that the birth/death rate $\mu$ vanishes.
This condition assures integrability of the system due to the existence of two functionally independent conserved quantities $q_1$ and $q_2$.
The integrable SIR model with vaccination and without birth/death (\ref{eq:SIRvO1}-\ref{eq:SIRvO3}) (equivalently (\ref{eq:SIRv1}-\ref{eq:SIRv2}) or \eqref{eq:2ndODE}) is simply abbreviated to the SIRv model.
If the vaccination rate $\nu$ moreover vanishes the SIRv model reduces to the original SIR model (\ref{eq:KMSIR1}-\ref{eq:KMSIR3}) preserving the integrability.
First we propose the discretization (\ref{eq:dSIR1}-\ref{eq:dSIR2}) of the original SIR model which possesses the same conserved quantities $q_1$ and $\tilde q_2=\left.q_2\right|_{\nu=0}$ as the continuous model.
Since the discretization, called the dSIR-I model, is moreover time-reversible, it is the integrable discretization of the SIR model. 
We also find a geometric construction of the dSIR-I model in terms of the non-autonomous parallel translation of the line possessing slope $-1$.
The intersection of this line and the invariant curve $\tilde\Gamma=\left.\Gamma\right|_{\nu=0}$ linearizes the time evolution of the dSIR-I model.
The trajectories of the SIR and  dSIR-I models form the same invariant curve $\tilde\Gamma$, since it is provided by their common conserved quantities $q_1$ and $\tilde q_2$. 
It follows the parametric solution to the SIR model via the solution to the dSIR-I model.

Next we consider the discretization of the SIRv model.
A naive generalization of the dSIR-I model ($\nu=0$) to $\nu>0$ does not possess the time-reversibility and hence the integrability, notwithstanding it has the same conserved quantities $q_1$ and $q_2$ as the SIRv model. 
Instead, we first consider the SIRv model with positive vaccination rate $\nu>0$ and vanishing recovery rate $\gamma=0$.
Then discretize it.
We moreover generalize the discrete model thus obtained to $\gamma>0$ and find the discrete SIRv (dSIRv) model (\ref{eq:dSIRv1}-\ref{eq:dSIRv2}).
The dSIRv model possesses not only the conserved quantities $q_1$ and $q_2$ but also the time-reversibility.
Hence the dSIRv model is the integrable discretization of the SIRv model. 
Although the dSIRv model is defined implicitly, it is explicitly expressed in terms of the Lambert W function.
Moreover, we find the parametric solution to the continuous SIRv model via the dSIRv model.
If we put $\nu=0$ in the dSIRv model, we obtain the integrable discretization (\ref{eq:dSIRII1}-\ref{eq:dSIRII2}) of the original SIR model different from the dSIR-I model (\ref{eq:dSIR1}-\ref{eq:dSIR2}), which we call the dSIR-II model. 
The two integrable discrete models provide different trajectories on the common invariant curve $\Gamma$ even for the same initial point.

As is mentioned in \S \ref{sec:ISIRM}, the SIRv model (\ref{eq:SIRv1}-\ref{eq:SIRv2}) we investigate is a restriction of the SIR model with vaccination (\ref{eq:SIRvG1}-\ref{eq:SIRvG3}) assuming the birth/death rate $\mu$ to vanish.
However, $\mu$ can be considered to be small considerably, since it stands for the death rate not caused by the disease.
Therefore, the integrable SIRv model closely approximates the non-restricted SIR model with vaccination.
Thus, it is natural to expect that the several integrable discrete models obtained in this paper play important roles in the study of the SIR model with vaccination and/or other variants of the SIR model. 

Both the continuous and the discrete SIRv models are considered to be the Lotka-Volterra system imposing special boundary conditions \cite{HT95,Bogoyavlenskij08}.
Through the Lax representation, one can construct the conserved quantities of the Lotka-Volterra system.
Hence, it looks that we can find the non-algebraic conserved quantity $q_2$ to the SIRv model via the Lax matrix $L$ of the Lotka-Volterra system. 
However, it is not so easy to deduce the conserved quantity $q_2$ from the eigenvalues of $L$, since $L$ is not reduced to a finite matrix even if we impose the boundary condition relevant to the SIRv model. 
The conserved quantity $q_2$ is given by a power series in the initial variables, hence $L$ must be an infinite matrix, whereas the Lotka-Volterra system reduces to the two-dimensional system by imposing the boundary condition.
We will discuss this subject from the viewpoint of symplectic geometry by using the 1-form $d\Psi$ \eqref{eq:tautologicalform} in a forthcoming paper.   

\begin{acknowledgments}
This work is partially supported by JSPS KAKENHI Grant No. 20K03692.
\end{acknowledgments}

~

{\flushleft\textbf{AUTHOR DECLARATIONS}}

{\flushleft\textbf{Conflict of Interest}}

The authors have no conflicts to disclose.

{\flushleft\textbf{Author Contributions}}

\textbf{Atsushi Nobe}: Investigation (lead); Methodology (lead); Writing -- original draft (lead).

{\flushleft\textbf{DATA AVAILABILITY}}

Data sharing is not applicable to this article as no new data were created or analyzed in this study.

\nocite{*}
\bibliography{Nobe220826}

\end{document}